\documentclass[12pt]{iopart}
\usepackage{graphicx} % Required for inserting images
\usepackage{xcolor}
\usepackage{amsfonts}
\usepackage{hhline}
\usepackage[linktocpage=true]{hyperref}
\usepackage[numbers,sort&compress]{natbib}
\hypersetup{
colorlinks=true,
citecolor=ultramarine,
linkcolor=cadmiumgreen,
urlcolor=indigo(dye),
pdfauthor={},
pdftitle={},
pdfsubject={}
}
        
\newcommand{\asharp}{A$^\sharp$}

\newcommand{\ba}{\begin{eqnarray}}
\newcommand{\ea}{\end{eqnarray}}
\newcommand{\be}{\begin{equation}}
\newcommand{\ee}{\end{equation}}

\newcommand{\Msun}{M_\odot}

\definecolor{grey}{rgb}{0.4,0.4,0.4}
\definecolor{dullmagenta}{rgb}{0.4,0,0.4}
\definecolor{darkblue}{rgb}{0,0,0.4}
\definecolor{midblue}{rgb}{0,0,0.5}
\definecolor{midred}{rgb}{0.5,0,0}
\definecolor{orange}{rgb}{1,0.5,0}
\definecolor{lightbrown}{rgb}{0.75,0.5,0.25}
\definecolor{tan}{cmyk}{0.14,0.42,0.56,0}
\definecolor{djunglegreen}{cmyk}{0.99,0,0.52,0}
\definecolor{lightgreen}{rgb}{0,1,0}
\definecolor{olivegreen}{cmyk}{0.64,0,0.95,0.40}
\definecolor{midgreen}{rgb}{0.0,0.675,0.0}
\definecolor{darkgreen}{rgb}{0,0.5,0}
\definecolor{ultramarine}{rgb}{0.07, 0.04, 0.56}
\definecolor{cadmiumgreen}{rgb}{0.0, 0.42, 0.24}
\definecolor{indigo(dye)}{rgb}{0.0, 0.25, 0.42}

\begin{document}

\title{Cosmography with next-generation gravitational wave detectors}

\author{Hsin-Yu Chen$^1$, Jose Mar\'ia Ezquiaga$^2$ and Ish Gupta$^3$}
\address{$^1$ Department of Physics, The University of Texas at Austin, 2515 Speedway, Austin, TX 78712, USA}
\address{$^2$ Niels Bohr International Academy, Niels Bohr Institute, Blegdamsvej 17, DK-2100 Copenhagen, Denmark}
\address{$^3$ Institute for Gravitation and the Cosmos, Department of Physics, Pennsylvania State University, University Park, PA 16802, USA}
\ead{hsinyu@austin.utexas.edu, jose.ezquiaga@nbi.ku.dk and ishgupta@psu.edu}
\vspace{10pt}
\begin{indented}
\item[]\today
\end{indented}

\begin{abstract}
 Advancements in cosmology through next-generation ground-based gravitational wave observatories will bring in a paradigm shift. We explore the pivotal role that gravitational-wave standard sirens will play in inferring cosmological parameters with next-generation observatories, not only achieving exquisite precision but also opening up unprecedented redshifts. We examine the merits and the systematic biases involved in gravitational-wave standard sirens utilizing binary black holes, binary neutron stars, and neutron star-black hole mergers. Further, we estimate the precision of bright sirens, golden dark sirens, and spectral sirens for these binary coalescences and compare the abilities of various next-generation observatories (\asharp, Cosmic Explorer, Einstein Telescope, and their possible networks). When combining different sirens, we find sub-percent precision over more than 10 billion years of cosmic evolution for the Hubble expansion rate $H(z)$. This work presents a broad view of opportunities to precisely measure the cosmic expansion rate, decipher the elusive dark energy and dark matter, and potentially discover new physics in the uncharted Universe with next-generation gravitational-wave detectors.
\end{abstract}

\section{Introduction}

\begin{figure}[t!]
\centering
\includegraphics[width = 0.5\textwidth]{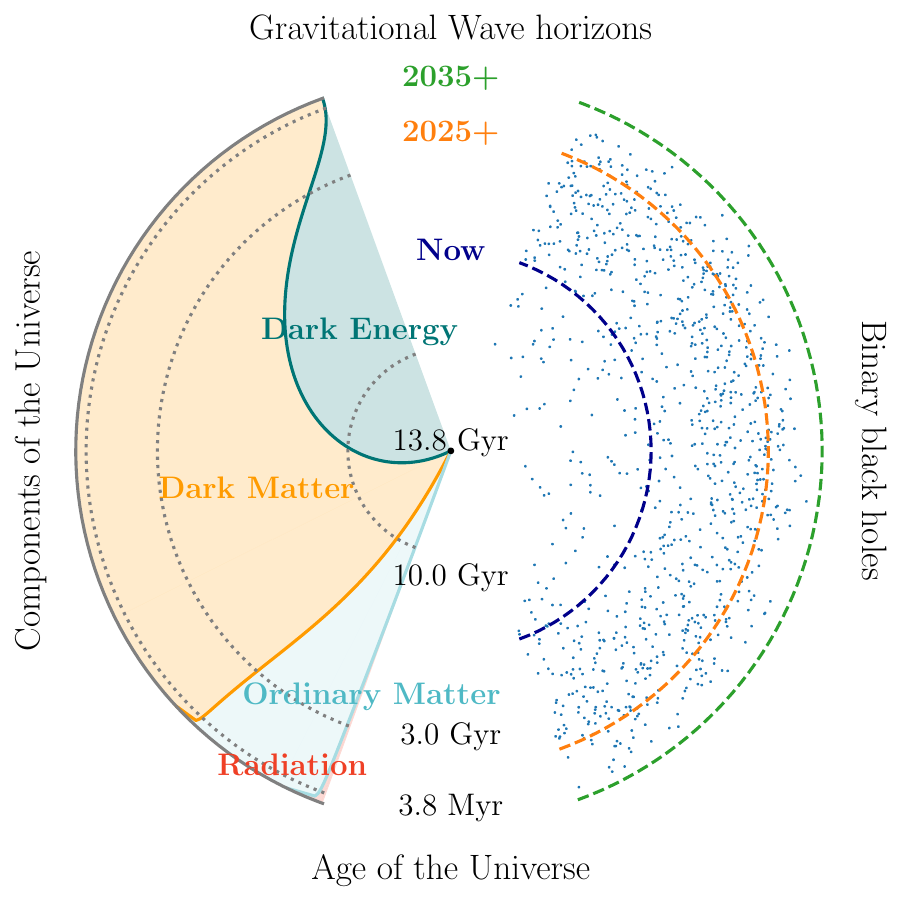}
\caption{Schematic diagram of the potential of GW observations to probe the evolution of the Universe. The radial coordinate represents the time from the present (center) to the beginning of the Universe (edge). 
On the right side of the diagram, distribution of stellar-mass BBH mergers (blue dots) and their associated horizon distance for different GW detectors (dashed lines). 
On the left side of the diagram, fraction of each Universe's component (dark energy, dark matter, ordinary matter and radiation) as a function of time, where a fraction of 1 at a given time is represented by the full arc.  
Although at present time we are observing a Universe dominated by dark energy, future GW detectors will observe an ancient Universe dominated by dark matter.}
\label{fig:gw_cosmography}
\end{figure}

In the next decades, gravitational-wave (GW) observations are expected to uncover many mysteries in the Universe. The planned next-generation ground-based and space-based GW observatories will bring a wealth of science opportunities with their reaches to the early Universe. 
The current generation of ground-based GW observatories, {LIGO \cite{Aasi2015}, Virgo \cite{Acernese_2014}, and KAGRA \cite{KAGRA:2020agh}}, with their successive upgrades (e.g., A+ {\cite{Miller:2014kma,KAGRA:2013rdx}} and \asharp {\cite{T2200287}}) will be able to detect stellar-mass binary black hole (BBH) mergers up to redshift $z\sim 10$~\cite{Gupta:2023lga}. Next-generation (XG) ground-based GW detectors, Cosmic Explorer {(CE)} {\cite{Evans:2021gyd}} and Einstein Telescope {(ET)} {\cite{et}}, aim at observing binary neutron star (BNS) and neutron star-black hole (NSBH) mergers up to $z\sim 10$ as well as all stellar-mass BBHs in the Universe ($z\sim 100$).\footnote{{NEMO is another possible future ground-based detector with sensitivity similar to A+ but ranging all the way to $\sim4$kHz \cite{Ackley:2020atn}. Therefore, NEMO is optimized to detect the (post-)merger of BNS. We do not discuss NEMO in this work.}}  
These observations of mergers across cosmic time will make XG detectors new powerful cosmological probes, mapping the Universe from the present dark energy domination to the past dark matter era, cf. figure \ref{fig:gw_cosmography}. 
{Future GW detectors in space will also bring relevant cosmological constraints \cite{LISACosmologyWorkingGroup:2022jok,tiango,TianQin:2015yph,Kawamura:2020pcg}, but we focus on ground-based detectors only here.}

In this era of precision cosmology with many well-established cosmological measurements, GW observations have several unique roles to play. First, GW observations will \textit{independently} measure the cosmological parameters, with the potential to resolve tensions among existing electromagnetic (EM) measurements~\cite{Abdalla:2022yfr}. 
Second, the detections across different redshifts allow for the mapping of the Universe's expansion history with a \textit{single probe} {over more than 10 billion years of cosmic evolution}. 
Third, 
XG detectors will verify the consistency between the EM and GW cosmological measurements and reveal (if any) underlying \textit{new physics}. {These observations will test gravity in \textit{new regimes}, reaching unprecedented strong fields and cosmic scales}.  

%-FIGURE: GW COSMOGRAPHY-
%------------

Assuming general relativity, the distances to a given binary can be determined from the GW signals released from the binary; meanwhile, multiple techniques are available to estimate the redshift of the binary. 
With the distance and redshift estimate, GW sources can be used to measure the Universe expansion rate $H(z)$ {and construct the GW Hubble diagram}. 
{In analogy to the use of supernovae Type Ia as standard candles for cosmology \cite{SupernovaSearchTeam:1998fmf,SupernovaCosmologyProject:1998vns}, this} is known as the `standard siren' method~\cite{Schutz:1986gp,2005ApJ...629...15H}. The observations of {the} BNS GW170817 and its EM counterparts enabled the first standard siren measurement~\cite{LIGOScientific:2017adf}{, opening a new era in multi-messenger astronomy (MMA)}. After the method was proposed in 1986~\cite{Schutz:1986gp}, new ideas on redshift acquirement, developments in data analysis, applications to GW data, and investigations of systematic uncertainties promised an exciting path forward for standard siren measurements. {In this article, we} explore further the prospects of the standard siren {cosmography} in the XG era.

In the following, we first review different techniques to obtain redshifts for standard sirens and the cosmology landscape with GW and EM facilities in 2035+. We then provide future projections to discuss the prospects of standard siren cosmography with XG detectors. 
{For concreteness, we focus on the forecast of the expansion rate of the Universe and its implications for measurements of the parameters of the standard cosmological models and constraints on its possible extensions. 
The science of future GW detector however extends further and includes, for example, cross-correlation with cosmic surveys \cite{Alonso:2020mva,Mukherjee:2019wfw}, gravitational lensing \cite{Xu:2021bfn} and stochastic backgrounds \cite{Renzini:2022alw}.} 

\section{Standard siren cosmology} \label{sec:stand_siren_cosmo}

Gravitational waves are natural messengers to probe the cosmic history.  
Unlike most other transients in the Universe, their signals can be well-understood from first principles, as general relativity has clear predictions for the waveform of a compact binary coalescence. 
{For example, at leading order, the frequency evolution is determined by a specific combination of the detected component masses of the binary known as the chirp mass $\mathcal{M}_c=(m_1m_2)^{3/5}/(m_1+m_2)^{1/5}$: $\dot f_\mathrm{gw}\propto \mathcal{M}_c^{5/3}f_\mathrm{gw}^{11/3}$ \cite{Maggiore:2007ulw}.} 
Moreover, GW detectors are directly sensitive to the waveform of the signal (as opposed to the intensity) allowing us to observe signals further away, since the GW amplitude scales with the inverse of the luminosity distance (and not the inverse square of the distance). 
Since GWs are coherently detected, the absolute value of the luminosity distance is directly measured{---giving GW sources the name of standard sirens.}  
The GW strain, $h_\mathrm{gw}$, thus contains information about the expansion rate through the luminosity distance
\begin{equation} \label{eq:luminosity_distance}
    h_\mathrm{gw}\propto {\mathcal{M}_c^{5/3}}/d_L\,, \qquad d_L=(1+z)\int_0^z\frac{c\mathrm{d}z'}{H(z')}\,,
\end{equation}
where we have assumed a flat background cosmology. Under that assumption and focusing on a Universe dominated by dark energy and dark matter {(radiation is only relevant in the early Universe)}, the Hubble parameter measuring the expansion rate can be written as
\begin{equation} \label{eq:Hubble_parameter}
    H(z)=H_0\sqrt{\Omega_\Lambda + \Omega_m(1+z)^3}\,,
\end{equation}
where $H_0$ is the {Hubble constant and denotes the} local expansion rate $H_0=H(z=0)$, and $\Omega_\Lambda$ and $\Omega_m$ are the fractional energy density of dark energy and dark matter that, in a flat Universe, satisfy $\Omega_\Lambda +\,\Omega_m = 1$. 
{The latest cosmological observations from the Planck satellite result in $H_0 = 67.66\pm0.42$km/s/Mpc, $\Omega_\Lambda=0.6889\pm0.0056$ and $\Omega_m = 0.3111\pm0.0056$ \cite{Planck:2018vyg}. 
Alternative cosmological models are also possible, potentially modifying both the background cosmology $H(z)$ and the scaling of the strain with the luminosity distance. We will discuss those in section \ref{subsec:test_lcdm}.}

GWs are unique compared to other transients because they keep an almost pristine record of the time of the merger due to the negligible interaction with the medium during propagation. In addition, they are particularly well suited for statistical population analyses since GW detector networks monitor the whole sky continuously and their selection effects can be understood in detail with injection studies \cite{Tiwari:2017ndi}. 

What GWs do not provide, however, is a direct account of the precise {epoch} when the {binary} merged. 
{Cosmic} times, or redshifts, together with distances allow the measurement of the Universe's expansion rate at different epochs. 
Therefore, the entire game of GW cosmology is to obtain redshift information. 
In the following we review different methods to achieve this goal. 

\subsection{Bright sirens} \label{subsec:bright_sirens}
One promising approach to obtaining the redshifts is to look for the EM counterpart of GW sources. With the finite number of GW observatories across the globe, the available observatory baselines limit the 2-dimensional resolutions to the sky location of the GW sources~\cite{Aasi:2013wya,2016ApJ...829L..15S,2017ApJ...840...88C,2016PhRvD..93b4013S}. However, the observations of EM counterparts could improve the sky localization of the GW sources and potentially lead to the identification of the host galaxies. The redshifts can then be measured from the spectra of the hosts. This is known as the \textit{bright siren} approach. 

Kilonova AT~2017gfo and GRB~170817A~\cite{LIGOScientific:2017ync,0817grb,swope,DECam1,2017ApJ...848L..24V,2017ApJ...850L...1L,2017Natur.551...75S,2017Natur.551...64A,2017Sci...358.1559K,2017ApJ...848L..14G} were the first EM counterparts found to be associated with a GW source{, GW170817 \cite{LIGOScientific:2017vwq}}. With the precise localization, the host galaxy NGC 4993 was identified and the redshift could be determined {($z\sim 0.009$)}. 
This event led to a Hubble constant measurement of $H_0=70^{+12.0}_{-8.0}$km/s/Mpc~\cite{LIGOScientific:2017adf}. {With Advanced LIGO-Virgo-KAGRA (LVK) sensitivities, the Hubble constant can be measured to $\sim 2\%$ when combining $\sim 50$ BNS bright sirens~\cite{samayah0,hsinyusiren,2019PhRvL.122f1105F,2018MNRAS.475.4133S}. This precision could shed light on the tension between different Hubble constant measurements~\cite{Abdalla:2022yfr}. Over last few years, a variety of other types of EM emissions were proposed theoretically~\cite{Pan:2022gwf,2018PhRvD..98l3002L,2019ApJ...873L..13D,2019PhRvD.100d3025P,2019ApJ...883L..19Z,2021ApJ...912L..18E,2022ApJ...927...56D,2022MNRAS.514.5385N,2022arXiv220808808P,2022arXiv220714435M,2019PhR...821....1P,2013PASJ...65L..12T,2018PASJ...70...39Y,2016PhRvD..94f4046L,2016ApJ...819L..21L,2016ApJ...822L...9M,2016ApJ...821L..18P,2016ApJ...827L..31Z,2016ApJ...822L...7W,2017arXiv170102492D,2019ApJ...884L..50M} or suggested observationally~\cite{2020PhRvL.124y1102G,2022arXiv220903363T,2022arXiv220913004G,JWST:2023jqa} to be associated with BNS, NSBH, and BBH mergers{, although they could not be confidently verified}. If confirmed, the applications of the bright siren approach will be further widened~\cite {2022MNRAS.513.2152C,2021ApJ...908L..34G,2020arXiv200914199M}.   }

Over the last few years, the systematic uncertainties associated with bright sirens have been widely studied, such as the systematics due to the GW instrumental calibration uncertainties~\cite{2022arXiv220403614H,2020CQGra..37v5008S,2021arXiv210700129S}, the EM observation selection effect~\cite{2020PhRvL.125t1301C,2023MNRAS.520....1G,2023arXiv230710402C}, the biased EM-inferred binary viewing angle~\cite{2020PhRvL.125t1301C}, the peculiar velocity of the hosts~\cite{2020MNRAS.492.3803H,2020MNRAS.495...90N,2021A&A...646A..65M}, and the GW instrumental non-stationary noise~\cite{2022PhRvD.106d3504M,LIGOScientific:2019hgc,LIGO:2021ppb,2021PhRvD.103l4061E,2022arXiv220212762K}. A comprehensive study of the systematics and the developments of the mitigation methods are critical to ensure the value of standard siren measurements in cosmology{, especially when dealing with the large and precise catalog of bright sirens that XG detectors will provide}.

\subsection{Dark sirens} \label{subsec:dark_sirens}

While the bright siren approach can lead to precise measurement of cosmological parameters, the expected number of systems with detectable EM counterparts is much smaller than those without EM signatures. In fact, since its inception, the LVK network has detected about a hundred mergers that were not associated with EM observations and only one, GW170817, for which the EM counterparts were observed. 
{This reflects the current selection biases that favor BBH detections over BNSs or NSBHs, as well as the difficulty to follow up distant events with poor localiztions.} 
Fortunately, several approaches can utilize the mergers with no EM counterparts to infer the cosmological parameters. For this reason, such mergers are also referred to as \textit{dark sirens}.

The various approaches that employ dark sirens for cosmological inference differ in their methodology for obtaining the corresponding redshift measurement that is associated with these events. In this section, we restrict ourselves to the approaches that use galaxy catalogs {to statistically infer the redshift of the sources}  
\cite{Schutz:1986gp}. 
{An alternative} technique relies on the fact that the GW sources as well as the galaxy population are expected to follow the large-scale structure. Hence, the cross-correlation between their spatial clustering can be used to infer cosmological properties \cite{Oguri:2016dgk,Bera:2020jhx,Mukherjee:2020hyn,Ghosh:2023ksl}. This {cross-correlation} technique was applied to eight well-localized BBH events from the Gravitational-Wave Transients Catalog-3 (GWTC-3) \cite{KAGRA:2021vkt} to get $H_0 = 68.2^{+26.0}_{-6.2}\,\,\mathrm{km}\,\mathrm{s}^{-1}\,\mathrm{Mpc}^{-1}$ \cite{Mukherjee:2022afz}. LVK detectors at design sensitivities, together with DESI \cite{DESI:2016fyo} and SPHEREx \cite{SPHEREx:2014bgr} data, have been shown to constrain $H_0$ to $\mathcal{O}(1)\%$ with $5$ years of observation \cite{Diaz:2021pem}.

{The \textit{statistical dark siren} method} exploits galaxy catalogs to identify the galaxies that lie within the localization volume associated with the GW signal (note that this localization volume is obtained purely from the GW detection \cite{Singer:2015ema}) and obtain their redshifts. Together with the probability distribution of $d_L$ inferred from the GW signal, the redshifts for each of these potential hosts give corresponding probability distributions on cosmological parameters. The uncertainty in the knowledge of the true host galaxy is accounted for by statistically averaging over the redshifts from all potential host galaxies, which results in constraints on cosmological parameters using a single GW event. Such an analysis can be performed for multiple GW events, which will result in the inference of cosmological parameters with greater precision \cite{hsinyusiren,DelPozzo:2011vcw,Gair:2022zsa,Borghi:2023opd}.

GW170814 \cite{LIGOScientific:2017ycc} was the first BBH merger to which this technique was applied. {It was localized to a volume of $2\times10^{6}\,\mathrm{Mpc}^3$ \cite{Palmese:2021mjm} which, according to the Dark Energy Survey (DES) data, contained $\sim77,000$ galaxies \cite{DES:2019ccw}. In comparison, the localization volume of the best localized BBH, GW190814 \cite{LIGOScientific:2020zkf}, was $\sim100$ times smaller and contained $\sim 2000$ galaxies \cite{DES:2020nay}.} Using GW170814 and the DES data \cite{DES:2018gui} for redshift information, $H_0$ was found to be $75^{+40}_{-32}\,\,\mathrm{km}\,\mathrm{s}^{-1}\,\mathrm{Mpc}^{-1}$ \cite{DES:2019ccw}. When applied to the eight best-localized events in GWTC-3, stronger constraints are obtained on $H_0$, measured to be $79.8^{+19.1}_{-12.8}\,\,\mathrm{km}\,\mathrm{s}^{-1}\,\mathrm{Mpc}^{-1}$ \cite{Palmese:2021mjm}. The statistical siren approach can be applied to bright siren events as well. However, in general, this method is expected to provide weaker constraints than the bright siren technique due to the uncertainty in host galaxy identification. Ref. \cite{LIGOScientific:2018gmd} applies the statistical siren strategy to GW170817 to obtain $H_0 = 76^{+48}_{-23}\,\,\mathrm{km}\,\mathrm{s}^{-1}\,\mathrm{Mpc}^{-1}$. But, as mentioned in section \ref{subsec:bright_sirens}, due to unique host galaxy identification made possible by EM counterpart detection, the bright siren method provides much stronger bounds on $H_0$. The constraints obtained from the statistical dark siren and the bright siren approaches were combined for events in GWTC-3 to estimate $H_0$ with $\sim 15\%$ uncertainty \cite{LIGOScientific:2021aug,Palmese:2021mjm}. Note that this measurement is dominated by the bright siren constraints from GW170817.

{The advent of XG observatories will see an improvement in the measurement precision of both luminosity distance and sky-area \cite{Borhanian:2022czq,Iacovelli:2022bbs,Gupta:2023evt,Branchesi:2023mws,Gupta:2023lga}. This will result in smaller localization volumes containing smaller number of potential hosts compared to currently detected events \cite{Muttoni:2023prw}.} Among these dark siren mergers, there may be some that are so well localized in the sky that only one potential host galaxy is found in that localization volume \cite{Chen:2016tys,Nishizawa:2016ood}. This would allow unique identification of the host galaxy and lead to a very precise estimation of cosmological parameters. Such mergers are referred to as \textit{golden dark sirens}. Asymmetric (i.e., the masses of the compact objects are appreciably different) and face-off (i.e., the inclination angle $(\iota)$ is not zero) systems are touted to be ideal candidates for such sirens \cite{Borhanian:2020vyr,Gupta:2022fwd}. This is due to the excitation of higher-order modes in the GW waveform for asymmetric and face-off systems \cite{Roy:2019phx,Mills:2020thr,Gupta:2024bqn}, leading to the breaking of the degeneracy between  $d_L-\iota$, which leads to the precise estimation of parameters \cite{VanDenBroeck:2006ar,Arun:2007hu,Ajith:2009fz,Graff:2015bba,Shaik:2019dym,Krishnendu:2021cyi}. Similar expectations hold for precessing systems {in which the spins are not (anti-)aligned and there could be precession of the orbital plane} \cite{Vitale:2018wlg,Krishnendu:2021cyi,Loutrel:2023boq}.  
{Because golden dark sirens are typically the loudest events, their measurement of the luminosity distance will be extremely precise, outperforming those of bright sirens.}

Several studies have looked at prospects of measuring cosmological parameters with XG GW observatories. Ref. \cite{Muttoni:2023prw} show that the application of the statistical siren approach to the binaries detected by a future network with 1 ET and 2 CE observatories will yield a $0.8\%$ bound on $H_0$ (with $90\%-$confidence) in one year of observation. However, they also point out that this bound is dominated by the detection of golden dark sirens. In fact, Ref. \cite{Borhanian:2020vyr} claim that just by using BBH golden dark sirens, a network with A+ sensitivities can ascertain $H_0$ to $\mathcal{O}(1\%)$ precision (with $68\%-$confidence) and an ET+2CE network can measure $H_0$ to $\mathcal{O}(0.1\%)$ (with $68\%-$confidence) in two years of observation. Ref. \cite{Gupta:2022fwd} finds that observation of NSBH mergers are not expected to provide any golden dark sirens unless the network contains at least 1 CE or ET. With an ET+2CE network, the constraints on $H_0$ from NSBH golden dark siren mergers can range from $\mathcal{O}(0.1\%)-\mathcal{O}(1\%)$ (with $68\%$-confidence) in 2 years of observation, based on the local merger rate of NSBH systems. Due to the ability of NSBH mergers to act as effective dark sirens and, potentially, bright sirens \cite{Kyutoku:2021icp,Feeney:2020kxk,Gupta:2022fwd} earns them the name \textit{grey sirens} \cite{Gupta:2022fwd}. 
{In section \ref{subsec:forecast_golden_dark_sirens} we present our own forecasts for dark siren cosmology.} 

As the galaxy catalogs are biased towards bright galaxies due to apparent magnitude thresholds, the dark siren method is also susceptible to the incompleteness of galaxy catalogs {that will increase  the statistical uncertainty of the measurements} \cite{hsinyusiren,Gray:2019ksv,Finke:2021aom,Gair:2022zsa}. Ref. \cite{Gray:2019ksv} finds that $H_0$ can be constrained to $\sim4.5\%$ (with $68\%-$confidence) with 249 BNS mergers (treated as dark sirens) in Advanced LIGO sensitivity, with a galaxy catalog that has only $50\%$ completeness up to a distance of 115 Mpc. For certain systems that are very well localized (golden or otherwise), targeted follow-up by telescopes to identify potential host galaxies can be a viable solution to tackle catalog incompleteness. 

The dark siren approach is also prone to systematic uncertainties. 
As this approach relies on the accurate estimation of $d_L$, it is significantly affected by the uncertainty introduced due to calibration errors, which can affect the amplitude measurement by as much as $7\%$ \cite{Sun:2020wke}. Fortunately, these errors are expected to reduce in the future. 
The redshift estimation can be affected by peculiar velocity measurements, especially for nearby hosts. As an example, the uncertainty in the peculiar velocity for the host of GW170817 (located at $z\sim0.01)$ was $150$ km/s, leading to a redshift uncertainty of about $5\%$ \cite{LIGOScientific:2017adf}. Particularly, this can influence the golden dark siren estimates, which, due to the stringent selection criteria, are expected to occur in the region $z \leq 0.1$ \cite{Borhanian:2020vyr,Gupta:2022fwd}. 
{There are also possible systematics associated to the assumption that binaries are tracers of certain cosmic structures, for example the stellar/halo mass, although these GW events themselves could also be turn into probes of the host galaxy properties \cite{Vijaykumar:2023bgs}. 
Moreover, if not included in the inference as in the spectral siren method described in the next section, the assumptions about the population properties are also important source of systematic errors \cite{Virgo:2021bbr}.}

\subsection{Spectral sirens}

Even in the absence of EM {counterparts or galaxy catalogs}, GWs \emph{alone} are able to constrain the cosmic expansion when studied collectively. 
The reason is that \emph{all} GWs signals are stretched by the expansion, redshifting its characteristic chirping. 
As a consequence, we can only detect the redshifted masses of the binary from the frequency evolution of the signal, i.e. the detected mass is   
\begin{equation}
    m_\mathrm{det}=(1+z)m\,,
\end{equation}
where $m$ is the mass in the source frame. 
Since all binaries at the same luminosity distance will be redshifted by the same amount, if a mass scale is found in the data of the mass spectrum or can be predicted by (astrophysics) theory, then this reference mass scale can be used to reconstruct the redshift at that distance. 
This finding of spectral features that carry redshift information can be thought in analogy to the absorption or emission lines of galaxy spectra. 
In this sense, a catalog of GWs is indeed a powerful set of \emph{spectral sirens}.

\begin{figure}[t!]
\centering
\includegraphics[width = 0.5\textwidth]{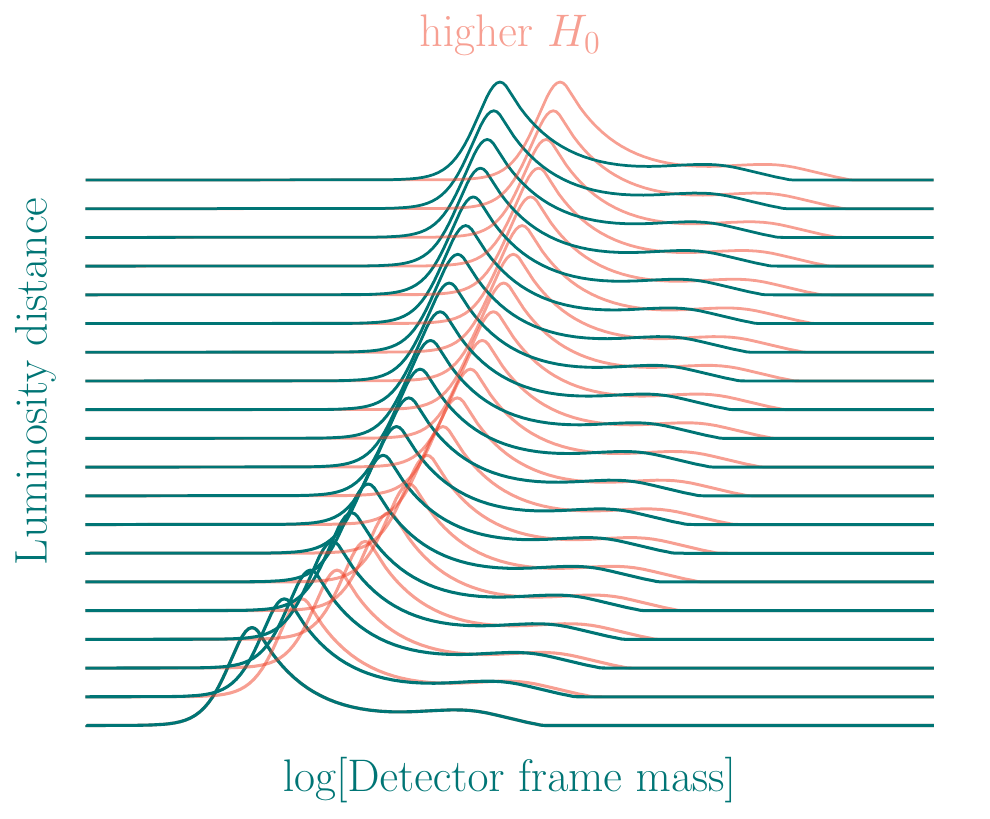}%,valign=t
\caption{Cartoon of the spectral siren method. For increasing luminosity distance, the whole distribution of detector frame masses shifts to higher values. The amount shifted corresponds to the redshift at a given luminosity distance and it is therefore sensitive to the expansion rate. For example, a higher value of $H_0$ associates higher detected masses to the same distance.}
\label{fig:spectral_siren_cartoon}
\end{figure}

For the same source population, different expansion rates will predict different distributions of detector frame masses. For instance, as schematically shown in figure \ref{fig:spectral_siren_cartoon}, a larger value of the local expansion rate $H_0$ will imply a larger redshift in order to end with the same luminosity distance, making the detector frame distribution shift faster to higher values. 
Therefore, one can constrain cosmological parameters with the population analysis of spectral sirens. 
Since the cosmological redshift is a common effect to all mergers at equal distance, the better the mass spectrum can be constrained over a larger range of distances, the more information can be extracted about cosmology. 
{One should be cautious, however, because the mass spectrum is expected to evolve with redshift and this could bias the cosmological inference if not properly accounted for, as we discuss later.}

%-FIGURE: CARTOON SPECTRAL SIREN-
%------------

The spectral siren method operates at its full capacity when there are clear features in the mass distribution.  
BNSs are expected to have cutoffs in their mass distribution from the existence of a maximum neutron star mass beyond which the star becomes a black hole and a minimum mass below which a white dwarf is formed. This narrow mass distribution made BNSs the first candidates for the spectral siren method~\cite{Chernoff:1993th,Taylor:2011fs}.  
Driven by observations, the mass distribution of BBHs have unveiled some interesting features. In particular there is a pronounced dearth of BBHs at high masses~\cite{Fishbach:2017zga, LIGOScientific:2021psn} and an excess over a simple power-law at $\sim35\Msun$ \cite{Farah:2023vsc}. At first sight, these features seem to coincide with the expectations from the theory of pair instability supernova (PISN)~\cite{Barkat:1967zz,Fowler:1964zz,Heger:2001cd,Fryer_2001,Heger_2003,2016A&A...594A..97B}, which predicts a gap in the BBH mass spectrum between $\sim50$--$120\Msun$ and a potential pile-up just before the lower edge of the gap \cite{Talbot:2018cva}. This connection has been however recently revisited \cite{Golomb:2023vxm}, and more observation are certainly needed to resolve the origin of these features. 
Irrespectively of their origin, these features stand out as clear targets for the BBH population to do spectral siren cosmology with current-generation detectors~\cite{Farr:2019twy} and have also been explored in the context of XG interferometers~\cite{You:2020wju}. 
With the latest LVK catalog, spectral siren constraints on $H_0$ are $\sim 20\%$ at $1\sigma$~\cite{Virgo:2021bbr}. 
{It is to be noted that if one \emph{assumes} some particular shape or location for these features, perhaps from some theoretical astrophysical prior, but those do not represent well the data, the cosmological parameters could be wrongly constrained.}

The spectral siren method is in its own nature data driven{: the cosmology is jointly inferred with the population properties which are directly obtained from the data}. It does not matter how the binaries are {astrophysically} formed or what they are made of, it only matters if a global shift of the mass spectrum can be identified over a wide range of distances {and distinguished from local changes of its shape}. 
In this sense all mergers can be considered simultaneously{, from light neutron stars to heavy black holes,} and cosmology is constrained with the full mass distribution of compact binaries \cite{Ezquiaga:2022zkx}. 
This is advantageous because different sectors of the mass spectrum are expected to evolve differently with redshift.  
For example, the masses of neutron stars and low mass black holes are thought to be much less sensitive to the environment {than} those of heavier black holes whose possible maximum mass depends significantly on metalicity.  
Therefore, by studying {the whole mass spectrum} together{, with its many individual features,} one can break the possible degeneracies between the shift induced by the cosmological redshift and the change in the shape of the mass spectrum due to the time and environment dependence of the source population \cite{Ezquiaga:2022zkx}. 
If unaccounted for, however, the evolution of the mass spectrum can bias the inference of cosmological parameters \cite{Ezquiaga:2022zkx,Mukherjee:2021rtw,Pierra:2023deu}{, a potential problem especially relevant for XG detectors observing far away binaries. 
Mismodeling of the true distribution with oversimplified parametric models \cite{Pierra:2023deu} or wrong astrophysical assumptions that do not fit well the data can also be problematic.}  
This is why the non-parametric reconstruction of the mass spectrum stand out as a promising avenue for spectral siren cosmology {with future observations, bypassing the need to choose a specific mass function and just relying on shape extracted from the (large) data sets \cite{Farah:2024xub}}.

{Finally,} it is important to note that our sensitivity to different sectors of the compact binary mass spectrum will evolve in parallel to the upgrades of the GW detector network. 
While currently the inference of cosmology is dominated by the excess of BBHs at $\sim35\Msun$ and the lack of them at $>45\Msun$, spectral siren cosmology with XG detectors will be dominated by BNSs and low-mass BBHs, as their intrinsic merger rate is much larger \cite{Ezquiaga:2022zkx}. 
Different populations will uncover the expansion rate at different redshifts, as their intrinsic merger rate history and detectability highly depends on their masses. 
Interestingly, with detectors at A+ sensitivity and more so with XG, we could also detect BBHs on the far-side side of the PISN gap (if they exist), thereby resolving the upper edge of the PISN gap and providing an extra anchor to probe cosmology~\cite{Ezquiaga:2020tns}. 

\subsection{Love sirens}

The ability to generate EM counterparts and the constrained mass spectrum make neutron star mergers effective bright and spectral siren candidates. %, respectively. 
But that is not all--- neutron stars that are part of compact binary coalescence (in either a BNS or an NSBH) undergo tidal deformation during the later stages of the inspiral. These deformations, at leading order, affect the phase evolution at the fifth post-Newtonian order (i.e, they are suppressed by a factor of $(v/c)^{10}$ compared to the dominant phase term, where $v$ is the orbital speed and $c$ is the speed of light) \cite{Hinderer:2009ca,Vines:2011ud}. The leading order tidal contribution is a linear function of the individual tidal deformability parameters of the two neutron stars, $\Lambda_1$ and $\Lambda_2$, and is called the reduced tidal deformability parameter $(\widetilde{\Lambda})$. If the neutron stars are spinning, the individual $\Lambda$s will also appear in the contributions to the phase from spin-induced quadrupolar deformations \cite{Nagar:2018zoe,Marsat:2014xea,Bohe:2015ana,Dietrich:2019kaq}. These changes in the phase allow the measurement of $\Lambda_1$ and $\Lambda_2$ when GWs are detected.  
Now, given an equation of state (EOS) for neutron stars, one can relate the individual $\Lambda$s to source-frame masses of the corresponding neutron stars. These source-frame estimates, along with the detector-frame redshifted masses estimated from the GW signal, can be used to ascertain the redshift associated with the binary \cite{Messenger:2011gi}. The post-merger signal of the BNS merger can also be used to break the redshift-mass degeneracy, by utilizing the robust spectral features that depend on the source-frame masses of the binary \cite{Messenger:2013fya}. These approaches that use the tidal deformability measurement and spectral features of the BNS system and the knowledge of the neutron star EOS to measure the redshift and obtain bounds on cosmological parameters constitute the \textit{Love siren} approach (as an ode to the neutron star tidal Love number \cite{Hinderer:2007mb}).

Note that, in practice, only $\widetilde{\Lambda}$ and not the individual $\Lambda$ are expected to be precisely measured, even with XG observatories \cite{Smith:2021bqc}.
However, there exist phenomenological relations, called quasi-universal relations \cite{Yagi:2015pkc}, that are independent of the EOS {governing the inspiralling neutron stars} and allow the estimation of individual $\Lambda$s using $\widetilde{\Lambda}$ and the mass-ratio of the binary. While this approach can indeed provide narrower bounds on the estimates of $\Lambda_1$ and $\Lambda_2$, they come at the cost of incurring biases due to uncertainties in the quasi-universal relations \cite{Yagi:2015pkc,Kashyap:2022wzr}. 

Using the inspiral technique for a BNS system in ET sensitivity, the redshift can be measured to a precision of $\mathcal{O}(10\%)$ for systems up to $z\sim1$ \cite{Messenger:2011gi,Li:2013via}. With the post-merger spectra, similar precision is obtained up to $z\sim0.04$ \cite{Messenger:2013fya}. Ref. \cite{Dhani:2022ulg} show that with the A+ network, less than $10$ detected BNS every year will be able to constrain the redshift to better than $10\%$ precision with the Love siren approach. However, with an ET+2CE network, the redshift can be resolved to $\sim1\%$ for $\mathcal{O}(10^4)$ yearly detections. This translates to $0.1\%$ error in $H_0$ and $0.61\%$ error in $\Omega_m$ with an ET+2CE network in a year of observation \cite{Dhani:2022ulg}. Ref.~\cite{Haster:2020sdh} shows that the XG observatories can constrain the EOS with BNS sources around $z\sim1$. 

The Love siren method provides a GW-only approach to the measurement of the cosmological parameters. It relies on the knowledge of the neutron star EOS, which is expected to be well determined by the XG era \cite{Finstad:2022oni,Huxford:2023qne,Gupta:2023lga,Shiralilou:2022urk}. 
So, either a previously inferred EOS can be used directly, or the EOS and the cosmological parameters can be jointly inferred for the GW data \cite{Ghosh:2022muc}. The forecasts that involve the post-merger spectrum require additional care, as they would strongly depend not only on the number of high-SNR BNS detections, but also on the post-merger modeling of the hypermassive neutron star remnant. Another aspect that requires caution, as noted earlier, is the use of quasi-universal relations to obtain individual $\Lambda$s from $\tilde{\Lambda}$ and the mass ratio. The fits for the universal relations themselves have uncertainties that can propagate to inference of cosmological parameters, inducing biases, especially at high precision. These uncertainties can be addressed by marginalizing over the residuals \cite{Chatziioannou:2018vzf,Kumar:2019xgp,Carson:2019rjx} or by specifically correcting for the incurred errors \cite{Kashyap:2022wzr}.

\section{The road to XG detectors} \label{sec:road_to_XG}

With more than one decade to go, the road to XG detectors will build up on many other efforts and will interplay with many other actors. 
On the GW side, several detector technologies still need to be demonstrated and their associated observing periods will continuously shed light on the cosmological origin of compact binary coalescences. 
On the EM side, standard siren cosmology will crucially depend on the capabilities of available telescopes to follow up the GW events. 
Moreover, cosmic surveys also aim at probing the cosmological model, acting as complementary partners to GW observatories.  
For this reason, before projecting the cosmology science case of XG detectors in section \ref{sec:new_horizons}, we briefly summarize the expected GW, EM, and cosmology landscape by 2035+ to understand better the new horizons opened by XG detectors. 

\subsection{GW astronomy in 2035+}
The detection of GW from compact binary mergers has opened a new window to the universe. The $\sim100$ events detected by the end of the LVK third observing run have resulted in constraints on deviations from general relativity \cite{LIGOScientific:2021sio}, inference of cosmological parameters \cite{LIGOScientific:2021aug}, revelations about neutron star properties \cite{LIGOScientific:2017ync,LIGOScientific:2017zic,LIGOScientific:2017pwl,LIGOScientific:2018cki,LIGOScientific:2018hze, LIGOScientific:2019eut} and measurement of the population properties of compact binary mergers \cite{KAGRA:2021duu}. However, based on the current estimates of the local merger rates and the redshift distribution of compact binary mergers, more than a million such events are expected to occur every year {within 600 Gpc $(z \sim 50)$} \cite{Gupta:2023lga}, most of which are missed by the GW networks at current sensitivities. Of the $\sim100$ events that are detected, the majority lie close to the SNR threshold of $10$, which does not allow the precise measurement of cosmological parameters, deviations from general relativity, or neutron star properties. Thus, to realize the full potential of GW, bigger and more sensitive detector networks are needed.

The first such advancement, called the A+ %(or O5)
concept \cite{Miller:2014kma,KAGRA:2013rdx}, is planned for the second half of 2020s.
With lower quantum noise and thermal coating noise, A+ is expected to be a $50\%$ improvement over advanced LIGO design sensitivity. 
In the most optimistic scenario, all five detectors, including the three LIGO detectors in Livingston, Hanford and the planned detector in Aundha (India), the Virgo detector in Italy, and the KAGRA detector in Japan, will have sensitivities similar to A+ by the end of 2020s. Apart from the improved sensitivity, such a 5-detector network would also excel in accurately pinpointing the location of mergers in the sky \cite{Borhanian:2022czq,Gupta:2023evt} and play an instrumental role in dark and bright siren cosmology \cite{hsinyusiren,Borhanian:2020vyr,Gupta:2023lga}. Beyond A+, one of the proposed improvements to the three LIGO detectors is the \asharp\ proposal \cite{T2200287} for the early 2030s. \asharp\ is expected to be about {twice} as sensitive as A+ across the frequency band. The improved sensitivity has a direct impact on the number of sources detected as well as the estimation of parameters, leading to more accurate distance and localization estimation, which will result in stronger constraints on cosmological parameters \cite{Gupta:2023lga}.

Even after the improvement in sensitivity, the science that can be explored with the A+ and the \asharp\ networks is limited by the size of the detectors. This is remedied by the proposed CE and ET observatories. The CE project \cite{LIGOScientific:2016wof,Reitze:2019iox,Evans:2021gyd,Evans:2023euw} involves the construction of two L-shaped detectors with arms of length 40 km for one detector and 20 km for the other. The order of magnitude increase in size and noise mitigation strategies result in $\mathcal{O}(10)-\mathcal{O}(100)$ improvement in the sensitivity compared to A+. 
On the other hand, ET \cite{Punturo:2010zz,Hild:2010id} is the proposed underground observatory in Europe, currently planned with three detectors with arms of length 10 km placed along the vertices of an equilateral triangle in a xylophone design (a comparison of different design proposals can be found in Ref. \cite{Branchesi:2023mws}, including the possibility of two underground L-shape detectors at different locations). ET is expected to have better sensitivity than CE at frequencies less than 10 Hz. Due to this, the inspiralling binaries remain in the sensitive frequency band of the observatory for much longer, which leads to an improvement in parameter estimation. Both the CE and the ET observatories are projected to begin observations in the late 2030s. 

The XG network with ET and the 2 CE detectors will be able to accomplish several science goals within just a few years of observation \cite{Branchesi:2023mws,Gupta:2023lga}. Not only will the majority of mergers occurring in the universe be detected, the measurement of source properties and localization will be performed at exquisite precision. As noted in section \ref{sec:stand_siren_cosmo}, this will enable GW observations to establish extremely stringent bounds on the measurement of the cosmological parameters, propelling our understanding of the universe to unprecedented levels.

%-FIGURE: GWs in 2035-
\begin{figure}[t!]
\centering
\includegraphics[width = 0.9\textwidth]{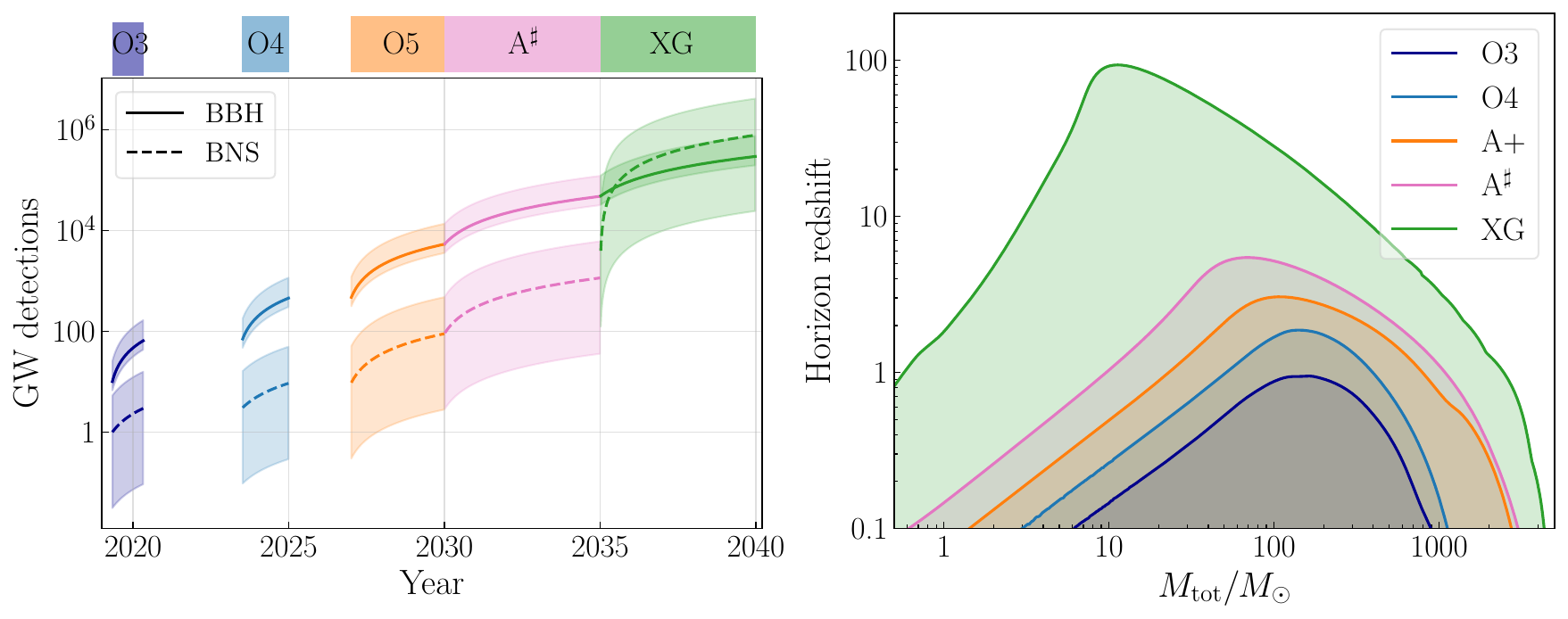}%,valign=t
\caption{On the left, expected, cumulative number of GW detections of BBHs (solid lines) and BNSs (dashed lines) as a function of time, taking into account the planned observing time and sensitivity of each proposed detector. 
The shaded bands represent the 90$\%$ confidence level on the local merger rate, which for BBHs ranges between $16\,\mathrm{Gpc}^{-3}\mathrm{yr}^{-1}$ and $61\,\mathrm{Gpc}^{-3}\mathrm{yr}^{-1}$ and for BNSs between $10\,\mathrm{Gpc}^{-3}\mathrm{yr}^{-1}$ and $1700\,\mathrm{Gpc}^{-3}\mathrm{yr}^{-1}$ \cite{LIGOScientific:2021psn}.
On the right, horizon redshift as a function of the total mass of the binary in solar masses. 
As a representative of XG detector's capabilities, we choose a 40km Cosmic Explorer.}
\label{fig:projection_gw_2035}
\end{figure}
%------------

To get a sense of how these planned improvements translate into numbers, in figure \ref{fig:projection_gw_2035} we project the expected number of detections as a function of year, taking into account our current understanding of the astrophysical population of compact binaries \cite{LIGOScientific:2021psn} and the current expectation for each detector to be operational. We split the count in BBHs and BNSs. 
From this graph it is clear that BBHs will continue to dominate the detection rates in the near future, while BNSs could potentially dominate in the XG era.
But, as emphasized before, it is not only that more events will be detected, each upgrade in sensitivity will open the possibility of exploring an uncharted redshift range. 
To exemplify this, we also present in figure \ref{fig:projection_gw_2035} the horizon redshift, i.e., the furthest a given GW detector can observe a signal, as a function of the total mass of the binary. 
As anticipated, XG detectors will reach all the way to redshifts of a hundred for BBHs and beyond the peak of star formation ($z\sim2$) for BNSs.\footnote{Note, however, that as the distance increases signals will be quieter in the detectors and, as a consequence, the inference of their properties will be degraded, including the distance itself \cite{Mancarella:2023ehn}.}

Apart from the ground-based observatories discussed above, the  Laser Interferometer Space Antenna (LISA) \cite{Babak:2021mhe}, a space-based GW observatory, is planned to launch in the mid-2030s. This will consist of three satellites 2.5 million kms apart, located at the vertices of an equilateral triangle, moving in a heliocentric orbit. This will allow LISA to observe GW in the mHz regime. While LISA will be able to constrain cosmological parameters to unprecedented precision by itself, the GW sources utilized for this science goal are distinct from the ones used by ground-based detectors \cite{LISACosmologyWorkingGroup:2022jok}. Instead of binaries containing neutron stars, massive BBH mergers are expected to serve as bright sirens. This is because the matter accreting on the massive black holes is expected to generate EM radiation close to merger \cite{Farris:2014zjo,Tang:2018rfm}. Despite the low expected detection rate, the exquisite sky localization \cite{Mangiagli:2020rwz} and the emission of EM radiation across all wavelengths \cite{DAscoli:2018dbt} render these sources as promising bright siren candidates. For dark siren measurements, stellar-mass BBHs within the redshift of $z=0.1$ can be employed as dark sirens, constraining $H_0$ to $\mathcal{O}(1)\%$ \cite{DelPozzo:2017kme}. Extreme-mass ratio {inspirals} at higher redshifs $(z\sim0.5)$ can also be used as dark sirens and can constrain $H_0$ to $\mathcal{O}(1)\%$ as well \cite{Laghi:2021pqk}. Moreover, as LISA can detect massive BBH sources to high redshifts $(z>1)$, it can utilize the effects of weak lensing to infer clustering parameters %like $\sigma_8$ and $n_S$ 
\cite{Congedo:2018wfn}. These sources can also be used to test $\Lambda$CDM cosmology and probe the nature of dark energy \cite{LISACosmologyWorkingGroup:2022jok}. LISA is also well positioned to test the validity of general relativity at cosmological scales \cite{LISACosmologyWorkingGroup:2019mwx}.

LISA will enable a direct interplay with XG detectors. Stellar-mass binaries could be seen by LISA during their early inspiral before merging in band of the XG network \cite{Sesana:2016ljz}. 
This is particularly interesting for a possible population of binaries in the far side of the PISN mass gap \cite{Ezquiaga:2020tns}, since the high-frequency sensitivity of LISA limits the detectability of binaries below the mass gap \cite{Moore:2019pke}.  
These ``multiband" detections will be good targets for cosmology as their first identification with LISA may allow for pre-merger localization, enhancing the possibility of finding multi-messenger counterparts. 
Even in the absence of counterpart the enlarged frequency evolution allow for a more precise localization in general that is beneficial for dark sirens. With just $\sim10$ detected events, $H_0$ could be constrained to a few percent \cite{Muttoni:2021veo}.

\subsection{EM observatories in 2035+} \label{subsec:EM_in_2035+}
Corresponding EM facilities will be critical for the GW siren measurements in the XG era. The observations of kilonova AT 2019gfo enabled the association to the host galaxy of {the} BNS GW170817, allowing for the first bright siren measurement~\cite{swope,LIGOScientific:2017adf}. Dark siren measurements with compact binary mergers observed by LVK also relied on the availability and completeness of galaxy catalogue~\cite{Virgo:2021bbr,Palmese:2021mjm,DES:2020nay}. 

\begin{figure}[t!]
\centering
\includegraphics[width = 0.9\textwidth]{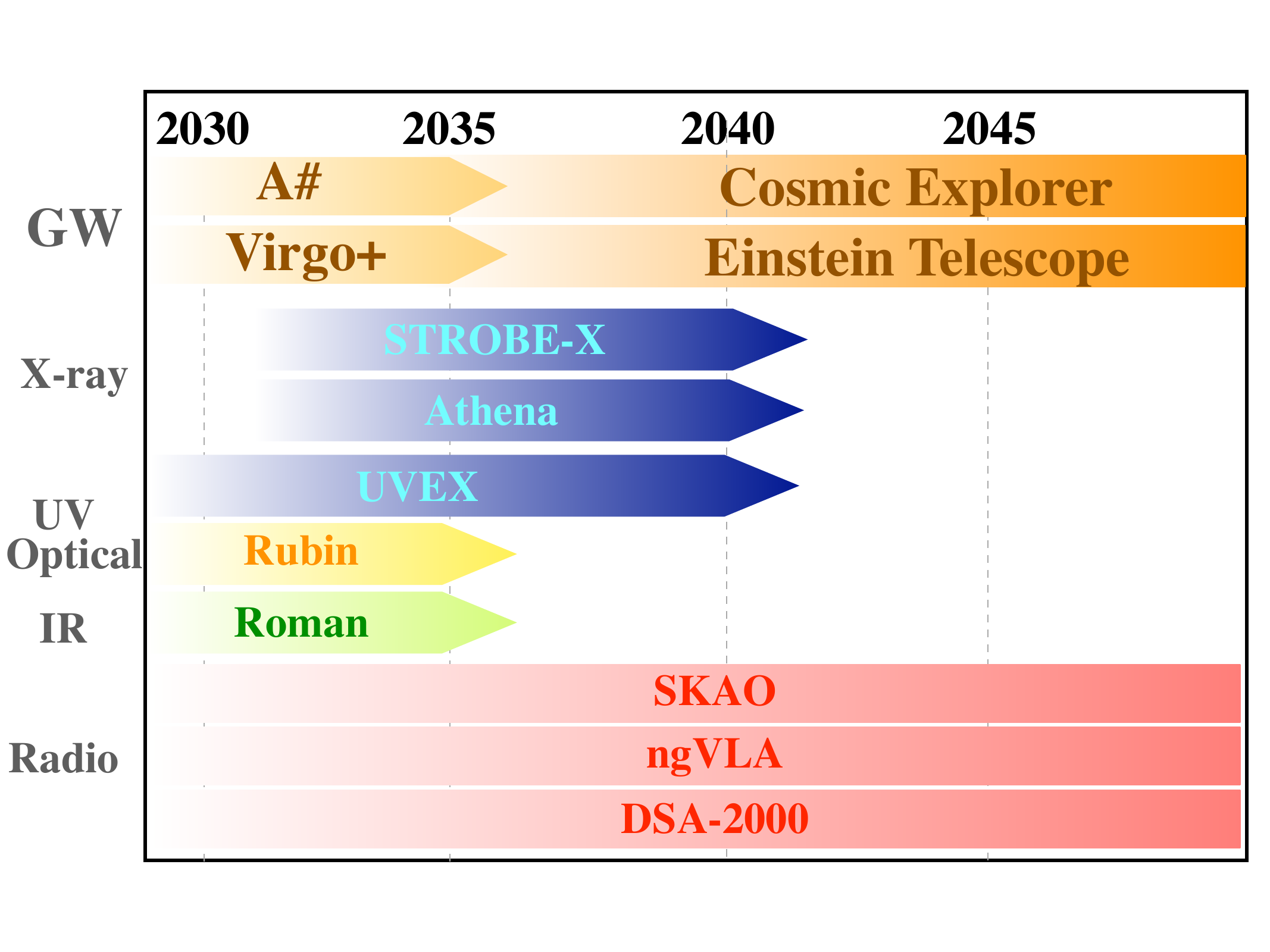}
\caption{Planned and proposed wide-field EM facilities that could contribute to the search for EM counterparts {in the XG era~\cite{2018SPIE10699E..19R,2021arXiv211115608K,2012arXiv1207.2745B,2015aska.confE.174B,LSSTScience:2009jmu,Spergel:2015sza,2022HEAD...1910846H,2018ASPC..517....3M,2019BAAS...51g.255H}. }
The \asharp\  network includes the two current LIGO sites, Hanford and Livingston in the US, and the planned Aundha in India.
}
\label{fig:projection_em_2035}
\end{figure}

EM facilities will be necessary to search for EM counterparts of GW events for bright sirens. Due to the large localization area by GW detectors ($\mathcal{O}(1 {\rm deg}^2)$ for BNS mergers at 500 Mpc for A\#, and $<2.5$ Gpc for XG ~\cite{Gupta:2023lga}) and the fast-fading nature of the EM counterparts (within weeks depending on the wavelength and emission mechanism), the search for EM counterparts will likely require instruments with the capability of wide-field coverage and prompt response. Given the large expected number of BNS and NSBH merger events ($\sim$ a few hundred a day), dedicated GW event follow-up instruments might be needed. In Figure~\ref{fig:projection_em_2035}, we present some of the planned and proposed wide-field telescopes across different wavelengths that could potentially contribute to the search for EM counterparts in the XG era\footnote{The figure does not mean to capture the comprehensive list of missions or to evaluate between missions.}.   

In addition to the search for EM counterparts, redshift measurement for the hosts of bright sirens and golden dark sirens will also require adequate EM resources. Especially, if high redshift EM counterparts are found, such as high-redshift GRBs, it is less likely to have existing galaxy catalog coverage and the follow-up observations will be necessary~\cite{2021ApJ...908L...4C}.

 Finally, deeper and more complete galaxy catalogs, such as those made available by Euclid~\cite{Amendola:2016saw} and DESI~\cite{DESI:2016fyo}, will play an important role in future dark siren measurements. If nearby bright sirens or golden dark sirens were found with peculiar velocity as part of the concern for systematic uncertainty, EM follow-up observations will also be useful for the reconstruction of the velocity field around the events if a complete galaxy catalog is not already available.

\subsection{Cosmology landscape in 2035+}

The study of the cosmos has seen a revolution in the last three decades with the discovery of the accelerated expansion of the Universe \cite{SupernovaSearchTeam:1998fmf,SupernovaCosmologyProject:1998vns} and the precise mapping of the anisotropies of the cosmic microwave background (CMB) \cite{WMAP:2003ivt,Planck:2018vyg}. These new cosmological probes have established a surprising ``standard" picture, called {the} $\Lambda$CDM model, in which most of the content of the present Universe is in the form of an unknown dark energy and dark matter. 
Much of the efforts of the large facilities that have been built and are planned to be built is precisely the unveiling of the physical origin of the dark Universe. 

The main cosmological probes are the observation of the primordial Universe with the CMB {\cite{WMAP:2003ivt,Planck:2018vyg}}, the measurement of the expansion rate with supernova type Ia (SNIa) {\cite{SupernovaSearchTeam:1998fmf,SupernovaCosmologyProject:1998vns}} and baryon acoustic oscillations (BAO) {\cite{SDSS:2005xqv}}, and the study of the growth of structure with galaxy surveys mapping the large scale structure (LSS) {\cite{DES:2021wwk}}, 
although there are many other emergent probes \cite{Moresco:2022phi}. 
All these probes aim at stress-testing $\Lambda$CDM in order to consolidate its position or discover new physics. 
Several tensions have been claimed over the years, but undoubtedly the most advertised one is the ``Hubble tension" \cite{Freedman:2017yms,Verde:2019ivm}, reporting a disagreement between the measurement of the local expansion rate  and the inferred value of $H_0$ from the CMB. 
It is unclear how long this tension will last, but, as we will see later, GW standard siren cosmology will provide a new way of precisely measuring the Universe's expansion rate.

There are currently multiple facilities with the potential to shed light on our understanding of dark energy and dark matter. 
Most notably, the Dark Energy Spectroscopic Instrument (DESI) \cite{DESI:2016fyo} and the Euclid satellite \cite{Amendola:2016saw} will be able to map millions of galaxies over 10 billion years of cosmic history ($z\sim 0$--2) to constrain the properties of dark energy and the growth rate of the LSS {using BAO, weak lensing maps and redshift space distortions among other methods. 
For example, DESI is expected to constrain the expansion rate $H(z)$ to a few percent precision over this wide range of redshifts \cite{DESI:2016fyo}. 
These surveys build} up on the efforts of their predecessors, the Dark Energy Survey (DES) \cite{DES:2021wwk} and the Sloan Digital Sky Survey (SDSS) \cite{SDSS:2005xqv}. 
In the near future, the Vera Rubin Observatory (VRO) \cite{LSST:2008ijt} will change the scale in which we map the Universe and observe its transient signals, playing a capital role in the multi-messenger transient endeavor. 
{For instance, VRO is expected to detect over 10 million supernovae during a 10-year period, which can be compared with the order 1500 that DES accumulated in 5 years \cite{DES:2024tys}.} 
The Nancy Grace Roman Space Telescope \cite{Spergel:2015sza} is expected to have a similar role from space, providing a detailed wide-field view of the sky. 
{Roman seeks to measure both expansion history and structure growth parameters with 0.1 – 0.5\% precision level, reaching 0.9\% and 2.1\% constraints on $H(z)$ between $z=$1-2 and $z=$2-3 respectively  \cite{Spergel:2015sza}.} 
More into the future, the Square Kilometer Array (SKA) \cite{Weltman:2018zrl} will look further back on time to explore the Universe when only the first galaxies were starting to be formed. 
{There are also proposals for next-generation spectroscopic surveys to follow DESI aiming at collecting more than an order of magnitude more galaxy redshifts \cite{DESI:2022lza}.}

\section{New horizons with XG detectors} \label{sec:new_horizons}

In this section, we will make projections on the measurement of cosmological parameters using GW observations. From the same, we have considered various detector networks, starting from \asharp\ sensitivity to a network with ET and two CE observatories. Details about the network configurations can be found in \ref{app:XGdetectors}. Specifically, we simulate populations of BBH, NSBH and BNS mergers consistent with current observations and following common astrophysical assumptions (details are described in \ref{app:mass_spin_dist} and \ref{app:redshift_dist}). The golden dark siren approach has been applied to the three compact binary classes and is presented in section \ref{subsec:forecast_golden_dark_sirens}. In section \ref{subsec:forecast_bright_sirens}, we generate kilonova light curves for BNS mergers and project the number of bright sirens based on the number of kilonovae that can be detected by LSST. The bounds on cosmological parameters using the spectral siren method with BBH and BNS mergers have been discussed in section \ref{subsec:forecast_spectral_sirens}. The joint forecast based on the estimates from these three sections has been discussed in section \ref{subsec:forecast_all_XG}.

\subsection{Golden dark sirens} \label{subsec:forecast_golden_dark_sirens}

Following the methodology presented in Refs. \cite{Borhanian:2020vyr, Gupta:2022fwd}, golden dark sirens are identified as mergers that lie within $z=0.1$ are can be localized in the sky to better than $0.04\,\mathrm{deg}^2$. Within $z=0.1$, only one $L^{*}$ galaxy \cite{1976ApJ...203..297S} is expected to be present in that small a sky-patch (Refs. \cite{Borhanian:2020vyr, Gupta:2022fwd} use equation (7) in Ref. \cite{Singer:2016eax} to arrive at this limit). As this would allow unique identification of the host-galaxy, we assume that the redshift is known for such systems. Then, the distance errors obtained from Fisher analysis using \texttt{gwbench} \cite{Borhanian:2020ypi} can be converted to constraints on $H_0$. Under the assumption of Flat $\Lambda$CDM cosmology, luminosity distance can be written a function of $H_0$ and the matter density, i.e., $D_L = D_L(\vec{\theta})$, where $\vec{\theta} = (H_0,\Omega_m)$. Then, the Fisher matrix obtained by combining the estimates from $N$ events is given as
\begin{equation} \label{eq:FIM_no_prior}
    \Gamma_{ij} = \sum_{k=1}^{N} \frac{1}{\sigma_{D_L}^2} 
    \left(\frac{\partial D_{L}}{\partial \theta_i}\right)\left(\frac{\partial D_{L}}{\partial \theta_j} \right)_k,
\end{equation}
where $\sigma_{D_L}^{2}$ is the error in luminosity distance from the GW detections obtained using \texttt{gwbench}. The square root of the diagonal elements of the covariance matrix $(\Gamma^{-1})$ associated with this Fisher matrix gives the constraints on $H_0$ and $\Omega_m$. 

In this current form, the Fisher matrix denoted in equation \ref{eq:FIM_no_prior} considers no prior information about the cosmological parameters. However, as $\Omega_m$ is not expected to be constrained by these nearby observations, the error on $\Omega_m$ estimated by the Fisher matrix will not only be unphysical (as $\Omega_m \in [0,1])$, but they will also adversely affect the errors in $H_0$. To mitigate this effect, we apply Gaussian priors to both $H_0$ and $\Omega_m$ with $(\sigma_{H_0},\sigma_{\Omega_m}) = (10,0.5)$. The prior on $H_0$ corresponds to the current bounds on $H_0$ from GW observations \cite{LIGOScientific:2021aug}. These can be included in the Fisher matrix calculation as \cite{Cutler:1994ys},
\begin{equation}
    \Gamma_{ii} = \sum_{k=1}^{N} \frac{1}{\sigma_{D_L}^2} 
    \left(\frac{\partial D_{L}}{\partial \theta_i}\right)^2_k\, + \frac{1}{\sigma_{\theta_i}^2}\,.
\end{equation}

\begin{figure}[h!]
\centering
\includegraphics[width = 1.1\textwidth]{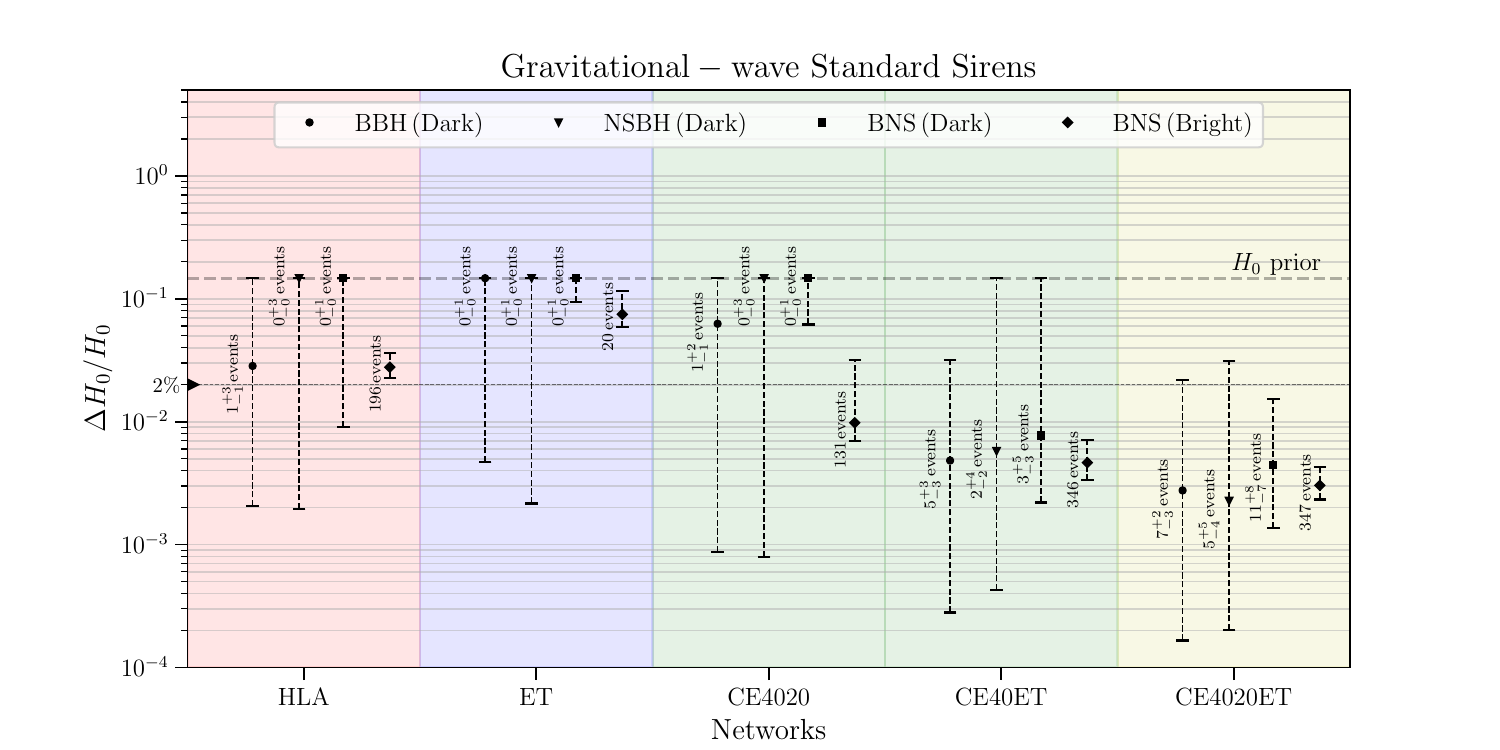}%,valign=t
\caption{Forecasts on the bounds that can be placed on $H_0$ with golden dark sirens and bright sirens in the observation span of a year. The markers show the median value of the constraints from the $1000$ realizations of the universe, and error bars mark the $99\%$ confidence region. The number of events that contribute to the corresponding bounds is also shown beside the error bars. The dotted line shows the $2\%$ precision mark in $H_0$ measurement, and the dashed line shows the current standard siren uncertainty on $H_0$ $(\sim15\%)$, which has been used as a prior in the analysis.}
\label{fig:golden_dark_and_bright_sirens}
\end{figure}

Based on the currently estimated redshift distribution of events, only $\sim10$ BBH mergers, $\sim20$ NSBH mergers and $\sim120$ BNS mergers are expected to occur within $z=0.1$ every year. 
To {avoid statistical fluctuations due to cosmic variance on} our $H_0$ measurement estimates by our choice of the parameters of these handfuls of events, we generate $1000$ permutations of the universe, in each of which the parameters for these events are randomly selected from our chosen population models. The fractional error in $H_0$ is estimated for each realization of the universe. The median values of these estimates along with the $99\%$-confidence regions are plotted in figure \ref{fig:golden_dark_and_bright_sirens}.

Note that not all realizations of the universe will have golden dark sirens. Thus, whenever a realization has no dark siren, we simply get back the prior that was put on $H_0$. From figure \ref{fig:golden_dark_and_bright_sirens}, we see that the number of golden dark sirens is much less compared to the bright sirens, but the constraints on $H_0$ are comparable for most cases. For the \asharp\ network, BBH golden dark sirens can {constrain} $H_0$ to $3\%$. Most of the realizations of the universe do not contain any NSBH or BNS golden dark sirens in the \asharp\ detections, leading to the median value coinciding with the $H_0$ prior. This is also the case with a network that contains just the ET observatory. However, here, even the BBH golden dark sirens are not expected to contribute significantly to $H_0$. This is due to the stringent sky-localization constraint on the selection of golden dark sirens, where a network of three \asharp\ detectors located far away from each other is seen to perform better than the three co-located detectors that make up the ET observatory. A similar observation is made for CE4020, where the two-detector network, with both detectors located in the US, is not able to measure the sky area well enough to qualify the detections as golden dark sirens. The CE40ET network does a much better job at constraining $H_0$, doing so to sub-percent precision for all the three classes of golden dark sirens. The constraints improve by a factor of 2-3 with the inclusion of CE20 in this network. Remarkably, for CE4020ET, all the realizations contain at least one golden dark siren, and almost all of them can constrain $H_0$ to better than $2\%$, resolving the $H_0$ tension in a year of observation.

\subsection{Bright sirens} \label{subsec:forecast_bright_sirens}

As seen in section \ref{subsec:bright_sirens}, BNS mergers can lead to a variety of EM counterparts which can be detected by various EM observatories (see section \ref{subsec:EM_in_2035+}). In this section, we focus on kilonovae that are detected by LSST for bright siren measurement. From the simulated population of detected BNS events, we choose the sub-population that lies within the redshift of $0.5$ and can be localized in the sky to $\Delta \Omega \leq 100\,\mathrm{deg}^2$, corresponding to each detector network. The choice of redshift corresponds to our expectation of the range of LSST for kilonova measurement, while the $\Delta \Omega$ cut-off ensures that LSST can observe the kilonova within $10$ sky patches. Then, using the kilonova model discussed in Refs. \cite{Villar:2017wcc,Kashyap:2019ypm,Wu:2021ibi}, we generate kilonova light curves in the $ugrizy$ bands of LSST, for all the events in these sub-populations. Assuming single-exposure with an exposure time of $30$ seconds, we claim that a kilonova is detected if its peak brightness exceeds the limiting magnitude of the telescope in that particular band. Using this approach, we find that most kilonova observations correspond to the $g-$band, with the maximum redshift at which a kilonova is detected to be $z=0.28$ (validating our $z\leq0.5$ cut-off for the sub-populations). Further, we assume that only $40\%$ of these kilonovae will \textit{actually} be detected \cite{Chen:2020zoq}. This fraction aims to account for the LSST sky coverage, systematic biases as well as duty-cycles corresponding to the GW and EM observatories. The remaining set of events, {those} that are detected both in the GW and the EM band, are used for $H_0$ measurement using the same methodology that was used for golden dark sirens in section \ref{subsec:forecast_golden_dark_sirens}.

The results for bright siren bounds on $H_0$ are portrayed in figure \ref{fig:golden_dark_and_bright_sirens}. We see that our bright siren estimates for $H_0$ are either at par with, or better than, the golden dark siren measurements. This disparity between the two approaches becomes notably significant for the HLA and the CE4020 networks. This is because HLA is not sensitive enough to resolve dark sirens to a very small area in the sky and uniquely identify the host galaxy. CE4020 is unable to achieve the same as it is only a two-detector network. The limitation of CE4020 to localize events in the sky is also evident in the bright siren estimates, where CE4020 detects 131 bright sirens, whereas the three-detector network HLA detects 196 such events. However, as the events detected by CE4020 have high SNRs and, accordingly, better distance estimates, the $H_0$ constraints with CE4020 are better than those from HLA.

We note that our bright siren estimates could be on the conservative side, as a result of our choice of the neutron star EOS, LSST detection strategy, or restricting ourselves to kilonova follow-up. {In particular, we chose APR4 \cite{Akmal:1998cf} to be the neutron star EOS, which results in relatively more compact neutron stars. Due to higher compactness, these neutron stars are difficult to disrupt tidally, reducing the amount of ejecta produced, which lowers the number of KN detections \cite{Gupta:2022fwd}).} Further, several studies have underscored the significance of short gamma-ray burst detections in enhancing the precision of bright siren measurements \cite{Chen:2020zoq,Wang:2021qwp,Dhani:2022ulg,deSouza:2021xtg}. These studies particularly emphasize their efficacy in probing events at high redshifts, thereby contributing to constraints on $\Omega_m$. Moreover, while our focus has been on kilonovae observations BNS mergers, a valuable avenue lies in considering NSBH mergers, which can also produce kilonova counterparts, thereby extending the pool of potential bright sirens for $H_0$ measurement \cite{Feeney:2020kxk,Gupta:2022fwd}. Nevertheless, amidst these considerations, the overarching conclusion is that BNS bright sirens can, by themselves, achieve $H_0$ measurements with sub-percent accuracy. This underscores the potential of BNS bright sirens as a powerful tool for precise cosmological parameter estimation.

\subsection{High-$z$ spectral sirens} \label{subsec:forecast_spectral_sirens}

Next-generation GW detectors will be revolutionary both in terms of the number of detections and redshift range. Compact binary coalescences will be detected every few minutes (approximately every 10 minutes for BBHs and every 3 minutes for BNSs), adding up to tens of thousands to hundreds of thousands of events per year. Of them, a large fraction will be beyond $z\sim1$. This makes XG CBC populations a perfect target for spectral siren cosmology, opening up the high-$z$ Universe.

We forecast the capabilities of spectral sirens performing a hierarchical Bayesian analysis on simulated data. 
We follow our fiducial population of BBHs and BNSs described in the appendices and simulate their detected posterior samples following the method described in \cite{Ezquiaga:2022zkx}. 
We assume a flat $\Lambda$CDM cosmology whose expansion rate is described by $H_0$ and $\Omega_m$, cf equation (\ref{eq:luminosity_distance}) and \ref{eq:Hubble_parameter}, taking as fiducial values the ones reported in Planck 2018 \cite{Planck:2018vyg}: $H_0 = 67.66$km/s/Mpc and $\Omega_m = 0.31$. 
We take a 40km CE sensitivity as representative for XG detectors. 
The inference is performed using \texttt{numpyro} \cite{phan2019composable} and accelerated with \texttt{JAX} \cite{jax2018github}.
Our code {is} available on \href{https://github.com/ezquiaga/spectral_sirens}{github}. 
The inference is performed across all the cosmological and mass function parameters. 
We do not consider NSBHs as their mass spectrum is still largely unknown. 

\begin{figure}[t!]
\centering
\includegraphics[width = 0.5\textwidth]{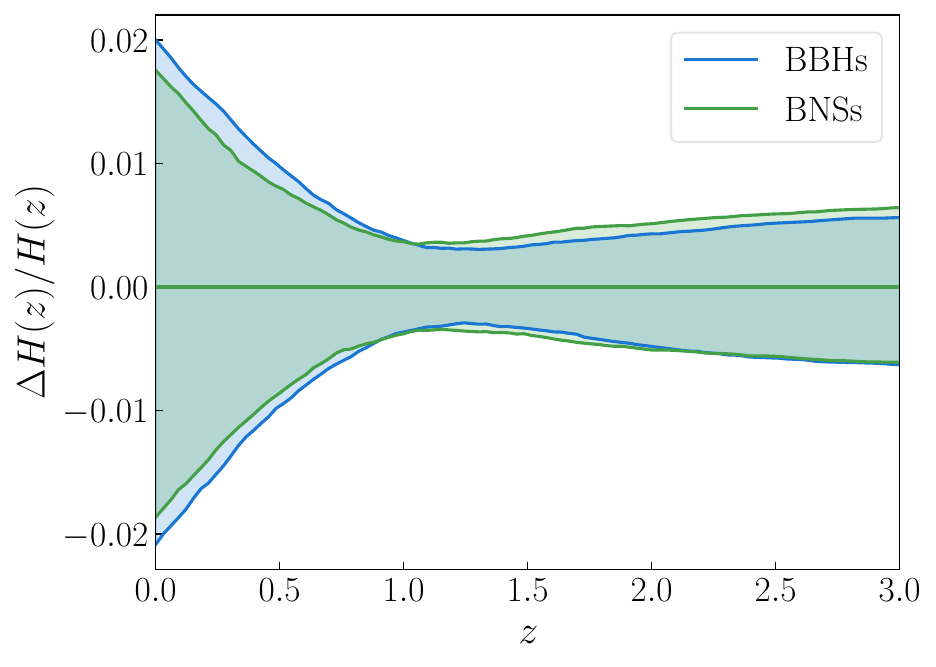}
\caption{Projected constraints on the Hubble parameter $H(z)$ with XG detectors. 
We compare the $1\sigma$ relative errors in $H(z)$ as a function of redshift for 50,000 BBHs and BNSs, using a 40km Cosmic Explorer detector as the fiducial sensitivity.}
\label{fig:spectral_sirens}
\end{figure}

In figure \ref{fig:spectral_sirens} we present the projected constraints on the expansion rate of the Universe as a function of redshift. 
We compare the $1\sigma$ relative errors for a fixed number of BBHs and BNSs, taking the error with respect to the mean of the $H(z)$ traces at each redshift.  
{Given current rate uncertainties,} we choose {a fixed number of} 50,000 events which corresponds approximately to the detection rate of BBHs per year with XG detectors (see figure \ref{fig:projection_gw_2035}). 
Since we assumed the same merger rate history $\mathcal{R}(z)$ for all compact binaries (see details in \ref{app:redshift_dist}) and BNS selection effects brings them to lower redshifts, the population of BNSs has more constraining power at lower $z$. 
Within the population of BBHs, it is the peak at low masses that dominates the inference \cite{Ezquiaga:2022zkx}. 
Both BNSs and BBHs can constrain the expansion rate with sub-percent precision over a wide range of redshifts, with the best constrained values at around $z\sim1$. 
{The best constrained redshift is located where most of the (well constrained) events are detected, which depends on the merger rate history and the detector's sensitivity.}
A combined analysis of all CBCs will only improve these constraints and allow to more efficiently break the possible degeneracies with the redshift evolution of the mass spectrum \cite{Ezquiaga:2022zkx}. 

%-FIGURE: SPECTRAL SIRENS-
%------------

\subsection{All-in XG cosmography} \label{subsec:forecast_all_XG}

Although we have introduced each type of standard siren separately for pedagogical reasons, they will all be analyzed simultaneously with XG detectors. In fact, as already discussed in the context of current detectors \cite{Virgo:2021bbr}, the population inference of the spectral siren method must be intertwined with the dark siren method, since {an incorrect population model} assumption can bias the dark siren cosmology. 
Recently, there have been developments of algorithms to jointly infer the GW population and cosmology with additional galaxy catalog information using hierarchical Bayesian analyses \cite{Mastrogiovanni:2023emh,Gray:2023wgj} and neural posterior estimation \cite{Leyde:2023iof}.

{To exemplify} a joint forecast of different standard sirens, we focus on {a population of BBHs and exploit} the capabilities of golden dark sirens to narrowly constrain the local expansion rate and on the exploration of the high-$z$ expansion history with spectral sirens. 
In practice, we take the projected constraints on $H_0$ from the Fisher analysis of golden dark sirens described in section \ref{subsec:forecast_golden_dark_sirens} and summarized in figure \ref{fig:golden_dark_and_bright_sirens}, and use them as the uncertainty for a Gaussian prior centered around the fiducial $H_0$ that is applied on the hierarchical Bayesian population analysis. 
As a benchmark point, we take an error in $H_0$ of $1\%$ to represent the capabilities of {BBH dark sirens with} a XG network. 

\begin{figure}[t!]
\centering
\includegraphics[width = 0.49\textwidth]{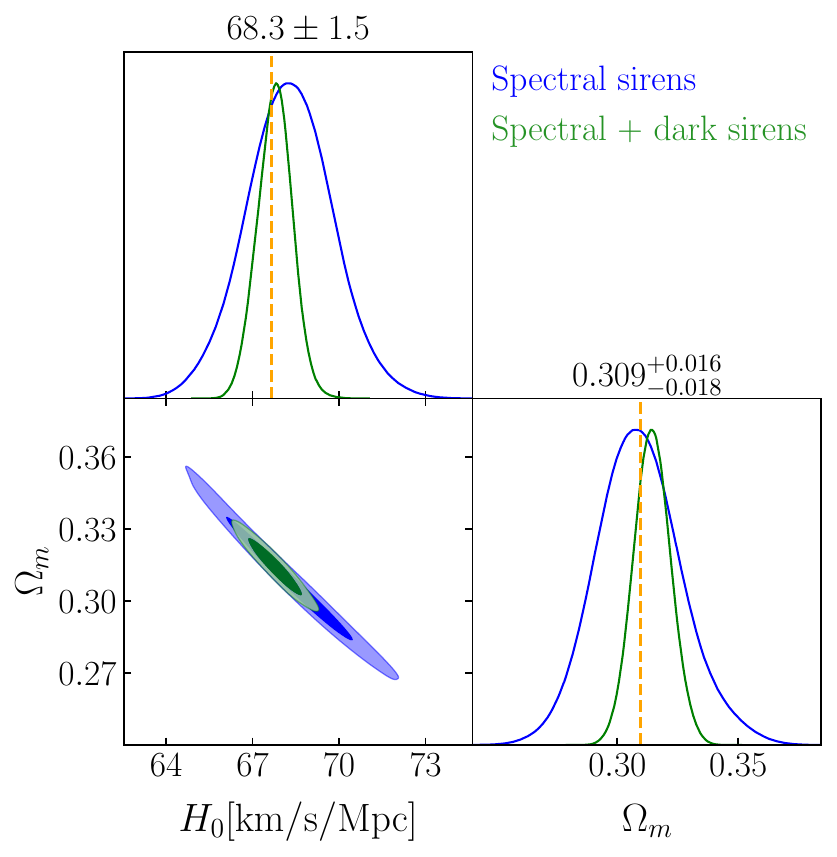}
\includegraphics[width = 0.49\textwidth]{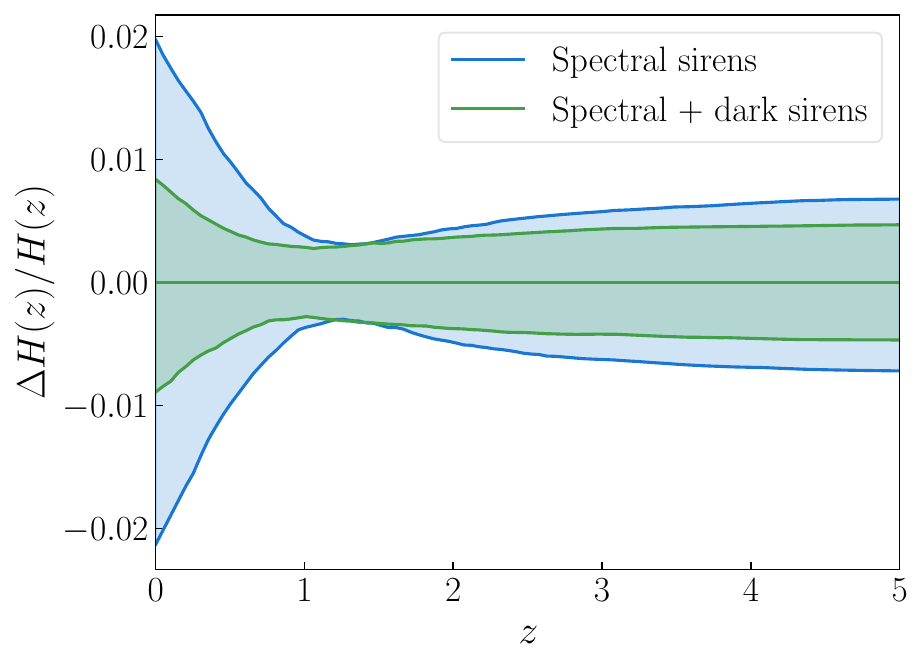}
\caption{Joint inference of spectral sirens and (golden) dark sirens with one year of observation of XG detectors. 
On the left, posterior distribution for {$H_0$} and $\Omega_m$. 
{The fiducial value is indicated with a dashed vertical line.} 
On the right, $1\sigma$ relative errors in $H(z)$. The dark siren prior in $H_0$ has a $1\%$ error and we use BBHs as the reference population for spectral sirens.}
\label{fig:spectral_sirens_joint}
\end{figure}

In figure \ref{fig:spectral_sirens_joint} we present the results for the joint forecast. 
We compare the combined analysis of spectral sirens and (golden) dark sirens with only spectral sirens {for a population of BBHs}. 
We show both the posterior distributions in {$H_0$} and $\Omega_m$ (left of figure \ref{fig:spectral_sirens_joint}) and the relative error in the expansion rare $H(z)$ (right of figure \ref{fig:spectral_sirens_joint}). 
As expected, the golden dark sirens dominate the local expansion rate while the spectral sirens are more powerful at $z>1$. 
Together, the expansion rate is constrained with sub-percent precision at all redshifts. 

%-FIGURE: SPECTRAL SIRENS - JOINT INFERENCE-
%------------

\subsection{Stress-testing the standard cosmological model}
\label{subsec:test_lcdm}

Although we have focused so far on the capabilities of future GW detectors to constrain the cosmological parameters describing the $\Lambda$CDM model, namely $H_0$ and $\Omega_m$ (see equation (\ref{eq:Hubble_parameter})), GW standard sirens are equally well suited to test deviations from this standard model. 
These generically come in two flavors: \emph{i)} as modifications in the properties of the different components of $\Lambda$CDM or \emph{ii)} as breaks in the foundational principles of the model.

For the first class, a lot of emphasis has been placed on postulating that the current accelerated expansion of the Universe is driven dynamically, for example, by a new scalar field. In practice this means that the energy density of dark energy is no longer constant, but rather evolves with time. This would change $\Omega_\Lambda$ in equation (\ref{eq:Hubble_parameter}) to $\Omega_\mathrm{de}(z)$.  
Since this dynamical evolution can in principle be quite arbitrary, phenomenological approaches in which the equation of state of DE is parametrized in simple terms have been common for encapsulating all the new physics, i.e. $\Omega_\mathrm{de}(z)=\Omega_\Lambda(1+z)^{3(1+w_\mathrm{de}(z))}$.  
Early work already anticipated that the equation of state of DE was a clear target for bright siren analyses with future GW detectors \cite{Sathyaprakash:2009xt}. 
Similarly, $w_\mathrm{de}$ was studied in the context of spectral sirens with the BNS mass function \cite{Taylor:2012db}.

For the second class, many (exotic) variants are possible. One that can particularly be well tested with GWs is that general relativity (GR) does not hold at cosmological scales. Among other things, this changes the propagation of GWs. 
Since compact binary coalescence occur at cosmological distances, even very small deviations in the GW propagation with respect to GR can be strongly constrained. 
The paradigmatic modifications of gravity at cosmological scales that have been explored are those that change the speed of propagation and the scaling of the luminosity distance with redshift \cite{Ezquiaga:2018btd}. 
In particular, for the case of scalar-tensor theories, the tremendous power of tests of the speed of gravity were anticipated \cite{Lombriser:2015sxa,Bettoni:2016mij} and spectacularly demonstrated with the first multi-messenger event, GW170817 \cite{Ezquiaga:2017ekz,Creminelli:2017sry,Baker:2017hug,Sakstein:2017xjx}. 
After the tight constraints on the speed of gravity{, which cannot differ from the speed of light in more than 1 part in $10^{15}$}, testing whether the GW luminosity distance is equal to the one of EM radiation has become the next goal. 
{This modified luminosity distance can be parametrized in terms of an additional friction term in the propagation equation $\nu$, where GR predicts $\nu=0$.}  
Current bright siren observations can only give weak constraints given the closeness of the sources \cite{Lagos:2019kds}{, $|\nu|< \mathcal{O}(30)$}, but spectral sirens can improve multi-messenger bounds given its larger redshift range \cite{Ezquiaga:2021ayr}{, $|\nu|< \mathcal{O}(3)$}. 
There has been a lot of activity to constrain this modified propagation with different standard siren methods \cite{Mastrogiovanni:2020gua,Finke:2021aom,Finke:2021eio,Chen:2023wpj,Mancarella:2021ecn,MaganaHernandez:2021zyc,Leyde:2022orh,Mukherjee:2020mha} and to forecast their future capabilities with XG detectors \cite{Belgacem:2019tbw,Maggiore:2019uih,Finke:2021eio,Balaudo:2022znx}{, which could improve by another order of magnitude}.  
Besides these effects, future GW detectors will be able to constrain other phenomena. For example, they will bound the presence of additional cosmological tensor fields \cite{Max:2017flc,BeltranJimenez:2019xxx}, which have a rich phenomenology of waveform distortions \cite{Ezquiaga:2021ler}, and test lensing effects beyond GR \cite{Ezquiaga:2020dao,Goyal:2023uvm}. 
Future GW observations combined with the redshifts from their host galaxies will also serve to measure peculiar velocities in order to constrain the growth of structure and test gravity \cite{Palmese:2020kxn}.

\section{Conclusions}

GW observations have opened a new window to explore the cosmos, unveiling a large population of BBHs and starting the era of multi-messenger astronomy with BNSs.
XG GW detectors will enter the ``big-data" phase and independently probe the Universe from low to high redshift. 
These advancements will be key in the quest for precision GW cosmology.

Different standard siren methods will complement each other in the XG era, having capabilities that cannot be achieved with current detectors. 
On the one hand, very loud, nearby events will have such a good localization that their EM counterparts can be easily sought for or only one galaxy may lay within their sky map, granting a precise measurement of the local expansion rate. 
On the other hand, the large population of compact binaries will allow to narrowly constrain the expansion rate at high redshift. 
When put together, we project that standard siren cosmography has the potential to constrain the Hubble expansion rate $H(z)$ to sub-percent precision over more than 10 billion years of cosmic evolution, {a precision that is at least comparable to (or better than) the projections from other direct measurements~\cite{2019astro2020T.270S} and some of the indirect measurements~\cite{2016arXiv161100036D,2022arXiv220903585S,2020PASA...37....7S,2021JCAP...12..049S}.}
This will be transcendental in multiple axes, having enough precision to arbitrate the tension among other cosmological measurements and to potentially discover new physics.  

The precise mapping of the Universe's expansion rate history with XG detectors will not only tightly measure the parameters of the standard cosmological model, but it will also allow to constrain its possible extensions. 
Dynamical dark energy is a good example, whose equation of state could be accurately bounded. 
XG detectors will also be in a perfect position to test gravity at cosmological scales and observe the lensing effects due to the inhomogeneities in the Universe.

The path toward such a powerful cosmological probe not only relies on the advancements in GW instruments but also the availability of corresponding EM facilities and a thorough understanding of the systematic uncertainties. The developments in experiments, theories, and data analyses in the next decade will be critical to pave the way toward the full strength of GW cosmography.

\ack
HYC is supported by the National Science Foundation under Grant PHY-2308752. JME is supported by the European Union’s Horizon 2020 research and innovation program under the Marie Sklodowska-Curie grant agreement No. 847523 INTERACTIONS, and by VILLUM FONDEN (grant no. 53101 and 37766). 
The Tycho supercomputer hosted at the SCIENCE HPC center at the University of Copenhagen was used for supporting this work. 
IG is supported by the NSF grants PHY-2207638, AST-2307147, PHY-2308886 and PHY-2309064. IG would also like to thank Rahul Kashyap for providing the scripts to generate kilonova light curves for BNS bright siren estimates, and Salvatore Vitale for the script to create the plot showing detector locations.

\appendix

\section{Detector locations and sensitivities}
\label{app:XGdetectors}
\begin{figure}[h!]
\centering
\includegraphics[width = 0.9\textwidth]{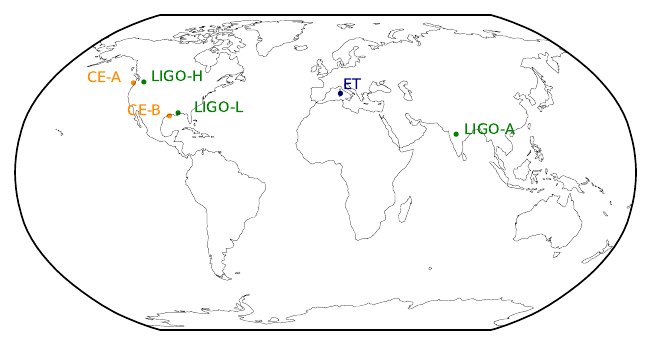}
\caption{Locations of the six ground-based GW detectors considered in this study.}
\label{fig:det_locs}
\end{figure}

\begin{figure}[h!]
\centering
\includegraphics[width = 0.7\textwidth]{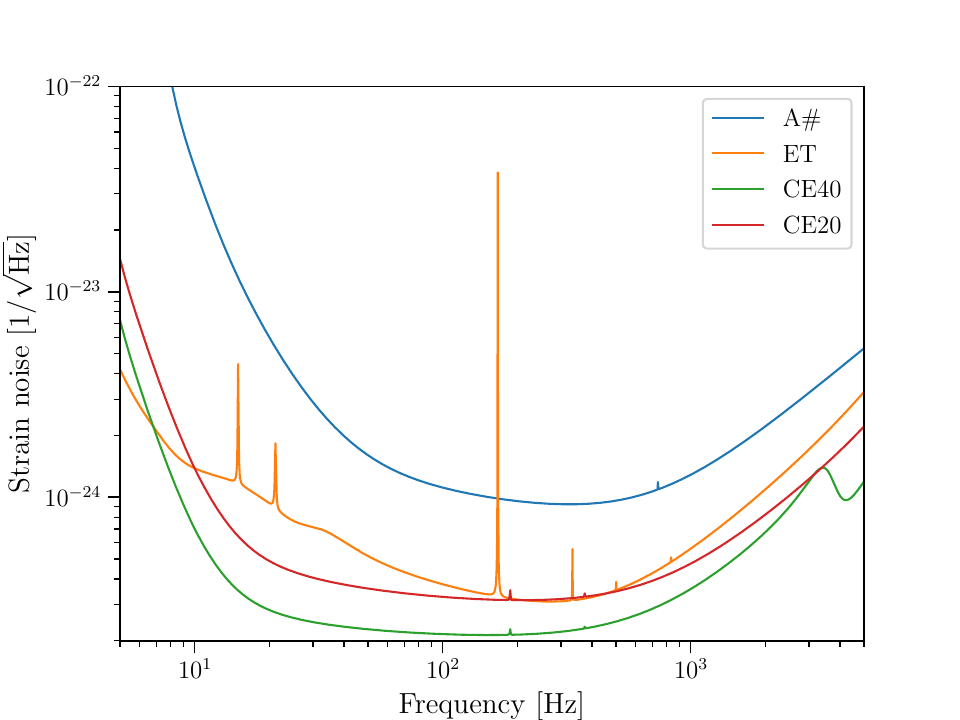}
\caption{The noise curves for the different detector sensitivities considered in this study.}
\label{fig:asd_noise_curves}
\end{figure}

The locations and the orientations of the GW detectors used in this study are mentioned in Table \ref{tab:locations} and shown in figure \ref{fig:det_locs}. The noise curves for the sensitivities that have been explored in this work are portrayed in figure \ref{fig:asd_noise_curves}.
\begin{table}[h]
\centering
    \begin{tabular}{l|l|l|l}
        \hhline{----}
        Detector & Latitude & Longitude & Orientation \\
        \hhline{----}
        LIGO-H & $46^\circ 27'18''$     & $-118^\circ 35'32''$     & $126.0^\circ $ \\ 
        LIGO-L & $30^\circ 33'46''$     & $-89^\circ 13'33''$      & $197.7^\circ $ \\ 
        LIGO-A & $19^\circ 36'47''$     & $+77^\circ 01'51''$      & $117.6^\circ $ \\
        CE-A & $46^\circ 00'00''$       & $-125^\circ 00'00''$     & $260.0^\circ $ \\ 
        CE-B & $29^\circ 00'00''$       & $-94^\circ 00'00''$      & $200.0^\circ $ \\
        ET   & $40^\circ 31'00''$       & $+9^\circ 25'00''$       & $90.0^\circ $ \\
        \hhline{----}
    \end{tabular}
    \caption{The position and the orientation of the detectors. Latitudes are positive in the northern hemisphere and longitudes are positive to the east of the Greenwich meridian. As all the detectors are in the northern hemisphere, all the latitude values are positive. The orientation is the angle north of the east of the $x$-arm. For L-shaped detectors, the $x$-arm is the one that completes a right-handed coordinate system together with the second arm and the local, outward, vertical direction. The x-arm of ET is defined such that the two other arms lay westward of it.}
    \label{tab:locations}
\end{table}

In sections \ref{subsec:forecast_golden_dark_sirens} and \ref{subsec:forecast_bright_sirens}, we consider five different detector networks and compare the bounds that can be put on $H_0$ with networks that contain these detectors. The names of these networks and the observatories that comprise them are given in table \ref{tab:net_names}. The HLA network contains LIGO-H, LIGO-L and LIGO-A at \asharp\ sensitivities. We did not consider Virgo and KAGRA for this network as it is still unclear if these two detectors will upgrade to similar sensitivities to \asharp, although their contributions at A+ level would still be very valuable for sky-localization estimates, aiding GW cosmology \cite{Gupta:2022fwd,Gupta:2023lga}. 
ET contains the triangular ET configuration, while CE40ET considers a network with both the 40 km CE detector (at CE-A location) and the triangular ET observatory. We also consider a CE-only network that contains the 40 km CE (at CE-A) and the 20 km CE (at CE-B). Finally, the most advanced network contains the 40 km CE, 20 km CE and the triangular ET.  
\begin{table}[!bt]
  \centering
  \begin{tabular}{l l l}
    \hline
    Network & Detectors in the network\\ 
    \hline
    HLA    & LIGO-L, LIGO-H, LIGO-A at \asharp\ sensitivity\\ 
%    \hline
    ET & Triangular ET  \\
%    \hline
    CE4020 & CE-A 40 km, CE-B 20 km \\ 
    CE40ET & CE-A 40 km, Triangular ET \\
%    \hline
    CE4020ET & CE-A 40 km, CE-B 20 km, Triangular ET \\
    \hline
\end{tabular}
\caption{Listed are the names of the five detector networks, along with the observatories that comprise these networks, that have been used in sections \ref{subsec:forecast_golden_dark_sirens} and \ref{subsec:forecast_bright_sirens}.}
\label{tab:net_names}
\end{table}

\section{Assumptions about the astrophysical population of compact binaries}

\subsection{Mass and spin distribution} \label{app:mass_spin_dist}
\begin{figure}[t!]
\centering
\includegraphics[width = 0.7\textwidth]{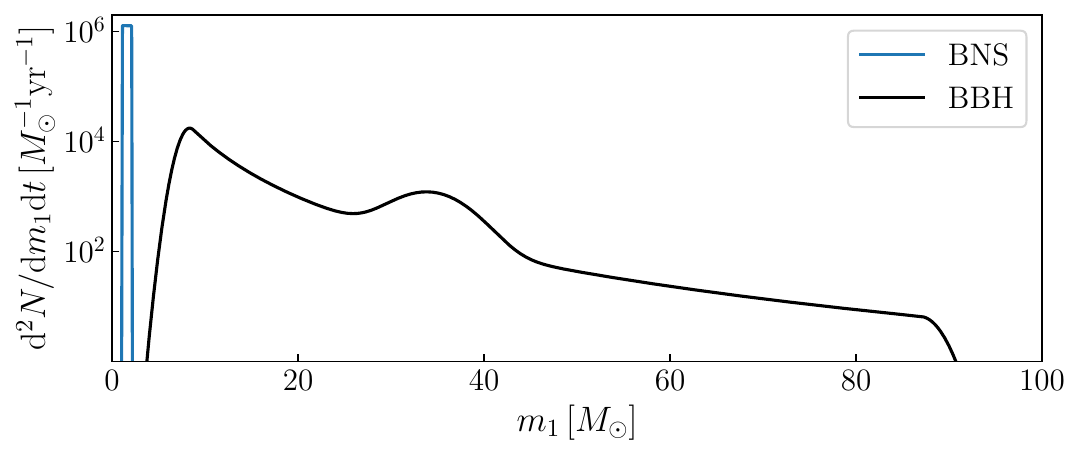}
\caption{Fiducial mass spectrum of compact binaries including BNSs and BBHs. This follows the observations from the latest GW catalog, GWTC-3 \cite{KAGRA:2021duu}.}
\label{fig:cbc_mass_spectrum}
\end{figure}

The population of BBHs is consistent with the inferred population with GWTC-3 events \cite{KAGRA:2021duu}. The population parameters for BNS and NSBH events are uncertain, owing to the lack of GW observations. The NS EOS is chosen to be APR4 \cite{Akmal:1998cf}. Following the maximum mass limit applied by APR4, we choose a uniform mass distribution between $[1,2.2]\Msun$ for neutron stars. {The fiducial mass spectrum of BNSs and BBHs is presented in figure \ref{fig:cbc_mass_spectrum}}.

We assume the population to have spins aligned with the orbital angular momentum. As precession, in general, is expected to improve the estimation of parameters \cite{Vecchio:2003tn}, the measurability estimates presented in this work may be conservative. The specifications for the three populations are given below:
\begin{enumerate}
\item{Binary black holes}
\begin{itemize}
    \item Primary mass: \texttt{POWER LAW + PEAK} \cite{KAGRA:2021duu} model with the following values for the model parameters: $\alpha=-3.4$, $m_{min} = 5\Msun$, $m_{max}=87\Msun$, $\lambda=0.04$, $\mu_{peak}=34\Msun$, $\sigma_{peak}=3.6$, $\delta_m=4.8\Msun.$
    \item Mass ratio: $p(q) \propto q^\beta$ with $\beta=1.1$, and enforcing $m_{min} = 5\Msun.$
    \item Spin magnitude: Aligned, independently and identically distributed (IID) spins following a beta distribution with $\alpha_\chi=2$, $\beta_\chi=5$ (see equation (10) in Ref. \cite{Wysocki:2018mpo}).
    \item Waveform: IMRPhenomXHM \cite{Garcia-Quiros:2020qpx}
\end{itemize}
\item{Binary neutron stars}
\begin{itemize}
    \item Mass: Uniform between $[1,2.2]\Msun$. 
    \item Spin magnitude: Aligned and uniform between $[-0.05,0.05]$.
    \item Equation of state: APR4
    \item Waveform: IMRPhenomPv2\_NRTidalv2 \cite{Dietrich:2019kaq}
\end{itemize}
\item{Neutron star-black holes}
\begin{itemize}
    \item Black hole mass: Same as the primary mass for BBHs.
    \item Neutron star mass: Same as mass for binary neutron stars.
    \item Black hole spin magnitude: Gaussian with $\mu = 0$ and $\sigma = 0.2$.
    \item Neutron star spin magnitude: Same as for binary neutron stars.
    \item Waveform: IMRPhenomXHM    
\end{itemize}
\end{enumerate}

%-FIGURE: MASS SPECTRUM-
%------------

\subsection{Redshift distribution} \label{app:redshift_dist}
\begin{figure}[t!]
\centering
\includegraphics[width = 0.7\textwidth]{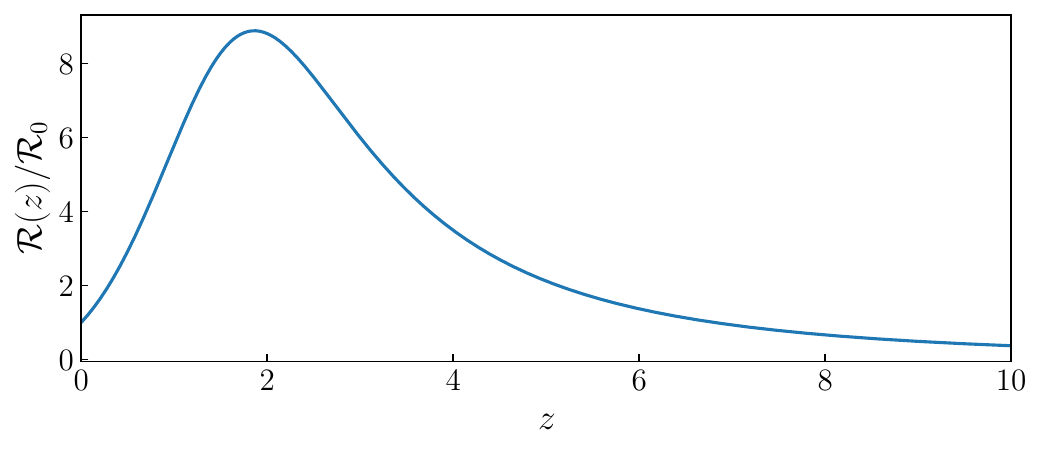}
\caption{Fiducial redshift distribution of compact binaries following the Madau-Dickinson star formation rate~\cite{Madau:2014bja,2019PhRvD.100d3030T}.}
\label{fig:cbc_redshift_distribution}
\end{figure}

For the three populations, we choose the merger rate to follow the Madau-Dickinson star formation rate~\cite{Madau:2014bja,2019PhRvD.100d3030T},
    \begin{equation}
        \psi(z|\gamma, \kappa, z_p) = \frac{(1 + z)^\gamma}{1 + (\frac{1 + z}{1 + z_p})^\kappa},
    \end{equation}
with $\gamma=2.7$, $z_{p}=1.9$, and $\kappa=5.6$. 
While the most accurate approach would be to convolve a chosen time-delay distribution with the star formation rate to obtain the merger rate, we have decided against choosing a time-delay distribution for this work. This is because the time-delay distribution is still uncertain, and the Madau-Dickinson star formation rate by itself is consistent with the current bounds on the merger rate with GWTC-3 events \cite{KAGRA:2021duu}. {Further, the redshift distribution is obtained by normalizing the star formation rate to match the local merger rate densities for the compact binaries.} Following the currently inferred rates \cite{LIGOScientific:2021qlt,KAGRA:2021duu}, the local merger rate densities for BBH, NSBH and BNS systems are chosen to be $\mathcal{R}_0^{\mathrm{BBH}} = 24\,\,\mathrm{Gpc}^{-3}\,\mathrm{yr}^{-1}$, $\mathcal{R}_0^{\mathrm{NSBH}} = 45\,\,\mathrm{Gpc}^{-3}\,\mathrm{yr}^{-1}$ and $\mathcal{R}_0^{\mathrm{BNS}} = 320\,\,\mathrm{Gpc}^{-3}\,\mathrm{yr}^{-1}$, respectively. The resultant redshift distribution is plotted in figure \ref{fig:cbc_redshift_distribution}.

%-FIGURE: REDSHIFT DISTRIBUTION-
%------------

\bibliographystyle{iopart-num}
\bibliography{bibliography_XGcosmography}
\end{document}